\documentclass[10pt]{IEEEtran}
\IEEEoverridecommandlockouts
\usepackage{cite}
\usepackage{amsmath,amssymb,amsfonts, amsthm}
\usepackage{graphicx}
\usepackage[table,cmyk]{xcolor}
\usepackage{hyperref}
\usepackage{multirow}
\usepackage{algorithm}
\usepackage{algpseudocodex}
\usepackage{caption}
\usepackage{makecell}
\usepackage{pgfplots}
\pgfplotsset{compat=1.18}

\author{S Ramprasath, Sibi Siddharthan, M G K S Charan, Vinita Vasudevan\\
    \IEEEauthorblockA{Department of Electrical Engineering\\Indian Institute of Technology Madras\\\{ramprasath, vinita\}@ee.iitm.ac.in, sibisiv.siddharthan@gmail.com,saicharanmarrivada0@gmail.com}
    }

\newcommand{\proc}[1]{\textnormal{\scshape#1}}
\newcommand{\thyphens}{\mathcode`\-=`\-\relax}
\newcommand{\func}[1]{\ensuremath{\mathop{\thyphens\operator@font#1}\nolimits}}
\newcommand{\id}[1]{\ensuremath{\mathit{\thyphens#1}}}

\newtheorem{theorem}{Theorem}

\newtheorem{lemma}[theorem]{Lemma}
\newcommand{\from}{\colon}
\newcommand{\B}{\mathbb{B}}

\usepackage{tikz}

\tikzstyle{vertex}=[circle,draw=black,minimum size=15pt,inner sep=0pt]
\tikzstyle{edgeu} = [draw,thick,-]

\begin{document}
\title{MCAC: A Model Counting Algorithm for Exact Computation of Error Metrics of Approximate Circuits}
\maketitle

\begin{abstract}
 Effective usage of approximate circuits for various performance trade-offs requires accurate computation of error. MCAC is a novel model counting framework for exact computation of several average and worst-case error metrics that are used to evaluate approximate circuits.
 Unlike other methods in the literature, our framework uses the same error miter for all metrics. It requires a single synthesis of the system consisting of  the exact and approximate circuits followed by a subtractor that finds the difference of the two outputs. Existing miter-based methods  require multiple calls to the model counter, one for each output of the miter.
 MCAC uses the  CNF formula of the system to compute all metrics. Our algorithm converts the  formula to a tree and uses message passing to compute all metrics. We propose data structures to efficiently store and perform sparse computations required for conversion to a tree and message passing.
Results for all the error metrics for several benchmark instances show a significant speedup over using off-the-shelf model counters along with specialized miters for each metric.
\end{abstract}

\section{Introduction}

Over the past decade, approximate circuits have gained traction as an effective method to trade off error for performance metrics like energy savings and frequency of operation in error tolerant applications. Computing the error in these circuits is an essential step towards determining the acceptability of the approximation. The system  used for error analysis consists of the exact and approximate circuits along with an error miter that models the desired error metric. In this paper, our focus is on exact computation of average and worst case error metrics that have been proposed in the literature. This includes standard metrics such as the error rate (ER), mean absolute error (MAE), mean squared error (MSE) and the worst case error (WCE) as well as the probability of worst case error (P(WCE)) and the entire probability mass function (PMF).

Exact evaluation of these error metrics is challenging, since the outputs of both the exact and approximate circuits are required for all possible values of the inputs.
In the past, methods used for exact error analysis include exhaustive enumeration, formal methods based on binary and algebraic decision diagrams (BDD/ADD) and model counting (\#SAT) \cite{Yu16,Vasicek19,Keszocze22,Mrazek17,Mrazek22,Meng24}, symbolic computer algebra (SCA) combined with ADD traversal \cite{Dreshler18,Dreshler19} and interactive theorem provers \cite{Qureshi19}.

Exhaustive enumeration is infeasible for larger circuits and/or circuits with a large number of inputs. Both SCA and theorem provers are limited to relatively small circuits. The most widely used techniques are the ones based on BDDs and \#SAT.  These methods are miter based and the entire system consisting of the exact and approximate circuits and the miter needs to be synthesized before error analysis. Each error metric requires a different miter and the computation of the metric requires a \#SAT call or a BDD for each output of the miter.
A significant overhead is the miter itself, which ranges from XOR gates that compare each output bit of the exact and approximate circuits to more complex circuits involving subtractors and comparators  \cite{Keszocze22,Vasicek19}.
Moreover, for a metric like MSE, the number of outputs of the miter can become very large (of the order of $n^2$ for $n$ erroneous outputs), making the computation impractical due to the large number of \#SAT calls/BDDs.

The complexity of error metric computation using BDDs and \#SAT depends on multiple parameters including  the number of outputs of the miter, the number of gates, the structure of the reconvergent loops in the fan-in cone  and the model-count of each output of the miter.
With BDDs, we have not seen results  for metrics like MAE and MSE for beyond 32-bit approximate adders and 12 bit multipliers \cite{Vasicek19}.
Computation of WCE alone is based on sequential calls to a SAT solver, which is possible for larger circuits. It has been used to design upto 32 bit approximate multipliers with guaranteed maximum error \cite{ceska17,Mrazek18,Mrazek18tvlsi}.
The method proposed in \cite{Meng24} use logic simulation within a SOTA model counter GANAK \cite{ganak} to speed up computation of the model count of each output of the miter. With tight coupling of the circuit structure and the CNF formula, they were able to compute the ER and MAE of 16 bit multipliers, but do not have results for either WCE or MSE.
The SCA based one-method-fits-all proposed in \cite{Dreshler19} is attractive, but has shown limited scalability.

\subsection{Contribution}
The two main bottlenecks in exact error metric computation are (a) Each metric requires a different miter and hence a separate synthesis of the system and (b) Computation of each metric requires the model counts at all outputs of the corresponding miter, making the process expensive. In this paper, we address both these issues.

We propose a model counting algorithm, MCAC, that uses a single error miter to compute  WCE, ER, MAE, MSE, P(WCE) as well as the PMF of the error.
It requires a single synthesis of the system  consisting of the exact and approximate circuits and a subtractor that finds the difference of the two outputs. This synthesized netlist is converted to a CNF formula ($F$) that is used by our algorithm to compute all metrics.

 Since the  fan-in cone of the miter outputs have significant overlap, it is inefficient to compute the model count of each output independently, as is done in all existing works.
MCAC is optimized to compute multiple model counts without starting from scratch each time.  To do this, we propose a novel algorithm that transforms $F$ to a tree  and uses a message passing algorithm  on the tree to compute all metrics. To improve computational efficiency, we propose data structures that exploit the sparsity of factors (satisfying solutions) and support the operations required for conflict removal and message passing.

We have tested the algorithm using instances from several benchmark sets in the literature, including 128 bit adders with as much as 90 error bits, 16 bit multipliers, and Gaussian blurring filters. For many of the instances, we have not seen any results of MSE computation using formal methods.

We get an average speedup of one to three orders of magnitude over making multiple calls to a SOTA model counter GANAK. For most instances, the runtimes are in the range of tens of seconds, making our algorithm suitable for use within an optimization framework.

\section{Related work}
In \cite{Dreshler18,Dreshler19}, symbolic computer algebra (SCA) is used to obtain the remainder of the approximate circuit with respect to specifications. Following this, they build and traverse an ADD for the remainder to get all metrics, including relative errors which is difficult to compute using a miter. The same method is used for all metrics, which is a significant advantage. They have demonstrated the method for upto 32 bit adders.

The miter based method to compute error metrics has been studied in considerable detail in \cite{Vasicek19}. A specialized miter is constructed for each error metric and the BUDDY \cite{buddy} is used to construct a (multi-output) BDD for each case. They also compare with model counting using SharpSAT  \cite{SharpSAT} as the counter. Other than relative errors, using BDDs they were able to compute metrics for upto 32 bit adders and 12 bit multipliers.  As with all miter based methods in the literature, a disadvantage is that each metric requires a separate miter. If several metrics are required, the synthesis of the system and analysis has to start from scratch using a new miter.
As reported in \cite{Keszocze22}, the time required to construct the BDD of the miter itself represents a significant percentage of the total computation time ($>$ 90\% for some outputs of small ISCAS85 benchmarks). The author optimizes the MAE miter for some special cases of the error and also proposes methods to compute the average and worst case relative error. The methods have been demonstrated for some outputs of ISCAS85 circuits and  32 bit adders. A more general simplification of the miter using ones-complement is presented in \cite{Mrazek22}. BDDs (CUDD \cite{cudd} and BUDDY) are used  to optimize upto 32 bit adders with the simplified miter. In \cite{ceska17} a SAT solver is used to compute the WCE of 32 bit multipliers. The miter used is a subtractor along with a threshold detector. The MAE is approximated using Monte Carlo analysis with  $10^9$ input vectors. In \cite{Mrazek17,Mrazek18}, the emphasis is on optimization of WCE, area and circuit metrics like power and delay to obtain a library of approximate adders and multipliers. Either Monte Carlo methods or exhaustive enumeration is used to obtain the average case metrics. In \cite{Mrazek25}, the WCE as well as the histogram of errors for approximate median filters is obtained by constructing BDDs for a  specialized sorting and comparison network. An SMT based technique is used in \cite{Rezaalipour23} to obtain the WCE, worst case Hamming distance and relative error.
In \cite{Meng24}, the authors propose VACSEM, in which the time required for each call to the model counter is optimized using logic simulation within the model counting framework of GANAK \cite{ganak}.
Here, the synthesized system is partitioned into multiple netlists, one for the fan-in cone  of each output of the miter.  With some significant synthesis effort using ABC \cite{Mishchenko06,abc} and tight coupling between the circuit structure and clauses in the CNF formula, they were able to push the boundary for computation of ER and MAE to 128 bit adders and 16 bit multipliers with a limited number of erroneous outputs.

Some works use a combination of formal methods and optimization to generate approximate circuits  that are guaranteed to have a particular error metric  less than a specified threshold. The method in \cite{Venkataramani12} is based on approximate logic synthesis, by identifying  approximation dont cares to simplify the function. A miter consisting of a subtractor and comparator for error thresholding is used along with either SAT, BDD or Monte Carlo anaylsis in \cite{Venkatesan11}. They have results for 32 bit adders and 8 bit multipliers. BDD minimization is used in \cite{Soeken16} to generate approximate circuits so that a given error metric is less than a specified threshold. The metrics used are WCE, ER and MAE and the method is demonstrated for some outputs of few of the smaller ISCAS85 benchmarks. In \cite{ceska22}, the authors demonstrate that a combination of error metrics is more useful for synthesis of approximate circuits. Their exploration is limited to 8 bit multipliers.

Besides GANAK and SharpSAT, other SOTA model counters include C2D and D4 \cite{C2D,D4}, SharpSAT-TD\cite{SharpSAT-TD} and GPU-SAT \cite{Fichte19}. To the best of our knowledge, none of them can efficiently  handle multiple queries of the kind required for error metric computation.

\section{Background}
\begin{figure}[tb]
\centering
\begin{minipage}[c]{0.3\columnwidth}
\centering
\begin{tabular}{|c|c|c|}
\hline
\multicolumn{3}{|c|}{$\phi_1(a, b)$} \\ \hline
$a$ & $b$ & $w$ \\ \hline
F & F & 0\\ \hline
F & T & 1\\ \hline
T & F & 1\\ \hline
T & T & 1\\ \hline
\end{tabular}\\
(a)
\\ \vspace{2mm}
\begin{tabular}{|c|c|c|}
\hline
\multicolumn{3}{|c|}{$\phi_2(b, c)$} \\ \hline
$b$ & $c$ & $w$\\ \hline
F & F & 1\\ \hline
F & T & 0\\ \hline
T & F & 1\\ \hline
T & T & 1\\ \hline
\end{tabular}\\
(b)
\end{minipage}
\begin{minipage}[c]{0.3\columnwidth}
\centering
\begin{tabular}{|c|c|c|c|}
\hline
\multicolumn{4}{|c|}{$\psi_3(a, b, c)$} \\ \hline
$a$ & $b$ & $c$ & $w$\\ \hline
F & F & F & 0\\ \hline
F & F & T & 0\\ \hline
F & T & F & 1\\ \hline
F & T & T & 1\\ \hline
T & F & F & 1\\ \hline
T & F & T & 0\\ \hline
T & T & F & 1\\ \hline
F & T & T & 1\\ \hline
\end{tabular}\\
(c)
\end{minipage}
\begin{minipage}[c]{0.3\columnwidth}
\centering
\begin{tabular}{|c|c|}
\hline
\multicolumn{2}{|c|}{$\psi_1(b)$} \\ \hline
$b$ & $w$ \\ \hline
F & 1\\ \hline
T & 2\\ \hline
\end{tabular}\\
(d)
\\ \vspace{2mm}
\begin{tabular}{|c|c|c|}
\hline
\multicolumn{3}{|c|}{$\psi_5(b,c)$} \\ \hline
$b$ & c & $w$ \\ \hline
F & F & 1\\ \hline
F & T & 0\\ \hline
T & F & 2\\ \hline
T & T & 2\\ \hline
\end{tabular}\\
(e)
\end{minipage}
\caption{(a), (b) Factor represented as a table for $\mathbf{F_1}(a,b)=a\lor b$, and $\mathbf{F_2}(b,c)=b\lor \lnot c$; (c) table for factor product $\psi_3(a,b,c)=\phi_1(a,b)\phi_2(b,c)$; (d) revisted table $\psi_1(b)$ after marginalizing $a$ in $\mathbf{F_1}$; (e) table for $\psi_4(b,c)=\psi_1(b)\phi_2(b,c)$} \label{tab:factor_ex}
\end{figure}
\subsection{The Message Passing Algorithm}
\label{sec:mp}
Message passing algorithms on trees are commonly used for Bayesian inference, see for example Chapter 10 of \cite{Koller2009} and\cite{pearl1982,Jensen90}.
In this section, we summarize some of the main definitions and results used in our algorithm.

Capital, bold letters (for eg. $\mathbf{X}$) are used to denote sets of variables. $|\mathbf{X}|$ is used to denote the number of elements in the set $\mathbf{X}$ and Domain($\mathbf{X}$) denotes the set of all possible assignments of variables in the set $\mathbf{X}$. $X$ denotes a particular assignment of variables. Small letters (for eg. $x$) are used to denote a single variable.

We use the following definitions in this section.\\
\noindent \textbf{Factor:} A factor $\phi(\mathbf{X})$ is a function
  that maps each assignment of variables in the set $\mathbf{X}$  to a non-negative real number, that is
  \begin{equation*}
    \phi(\mathbf{X}) : \textrm{Domain}(\mathbf{X}) \to \mathbb{R} \ge 0
    \end{equation*}
  In this paper, we represent the factor as a table $\mathbf{T(X)}$ with entries $\{X_i,w(X_i)\}$, $X_i \in  \textrm{Domain}(\mathbf{X}) \rightarrow w(X_i) \ge 0$. Consider an example Boolean function $\mathbf{F_1}=a\lor b$. Let $\mathbf{X} = (a, b) \in \mathbb{B}^2$, $\phi$ be a factor that has $w(X_i)$ as 1 if $X_i$ is a satisfying solution of $\mathbf{F_1}$ and 0 otherwise. $\mathbf{T(X)}$ representing the factor $\phi(\mathbf{X})$ is shown below in Table~\ref{tab:factor_ex}(a).\\
 \textbf{Partition Function (PF):} Let $\phi(\mathbf{X})$ be a factor and $\mathbf{T(X)}$, the corresponding table.  The PF of $\phi$ is defined as
  \begin{equation}
    \textrm{PF}(\phi) =  \textrm{PF}(\mathbf{T}) = \sum_{X \in \textrm{Domain}(\mathbf{X})}  \phi(X) = \sum_{X \in \textrm{Domain}(\mathbf{X})}  w(X)
    \label{eq:pf}
  \end{equation}
  For the factor in Table~\ref{tab:factor_ex}, the PF($\mathbf{T}$) is the total number of satisfying solutions of $\mathbf{F_1}$. \\
\textbf{Marginalization:}
  Let  $\phi(\mathbf{X},z), z \notin \mathbf{X}$ be a factor. $\psi(\mathbf{X})$ is a factor obtained from $\phi$ after marginalization of the variable $z$, if for each assignment $X_i \in$ Domain($ \mathbf{X})$
  \begin{equation}
    \psi(X_i) = \sum_{z \in \{0,1\}}\phi(X_i, z)
    \label{eq:marg}
  \end{equation}
  If the entries for $z = 0$ and $1$  in the table $\mathbf{T}(\mathbf{X},z)$ are $\{X_i,0,w(X_i,0)\}$ and $\{X_i,1,w(X_i,1)\}$ respectively, the entry in the revised table after marginalization of $z$  is $\{X_i,w(X_i,0) + w(X_i,1)\}$. Table~\ref{tab:factor_ex}(d) shows the revised table $\psi_1(b)$ after marginalization of $a$ in $\phi_1(a,b)$(Table~\ref{tab:factor_ex}(a)).

\textbf{Factor product:}
  Let $\phi_1(\mathbf{X}, \mathbf{Y}), \phi_2(\mathbf{Y,Z})$ be two factors. The factor product $\phi_1 \phi_2$ gives a factor $\psi$ which is obtained as follows.
\begin{align}
\forall X,Y,Z\in \textrm{Domain}(\mathbf{X,Y,Z}), \nonumber \\
\psi(\mathbf{X}, \mathbf{Y}, \mathbf{Z}=X,Y,Z) = \phi_1(X,Y) \phi_2(Y,Z)
\label{eq:fp}
\end{align}
For each assignment $Y_k$ such that  $\{X_i, Y_k, w(X_i,Y_k)\} \in \mathbf{T}_1(\mathbf{X}, \mathbf{Y})$ and $\{ Y_k, Z_j, w(Y_k,Z_j)\} \in \mathbf{T}_2(\mathbf{Y}, \mathbf{Z})$, the factor product of the two tables, $\mathbf{T(X,Y,Z)}$, has an entry $\{X_i, Y_k, Z_j, w(X_i,Y_k)w(Y_k,Z_j)\}$. The resultant PF can be written as
\begin{equation}
  \textrm{PF}(\psi) = \sum_{Y_k \in \textrm{Domain}(\mathbf{Y})} \sum_{\substack{X_i \in \textrm{Domain}(\mathbf{X})\\Z_j \in \textrm{Domain}(\mathbf{Z})}} w(X_i,Y_k)w(Y_k,Z_j)
\end{equation} An example factor product, $\psi_3(a,b,c)=\phi_i(a, b)\phi_2(b,c)$, for the function $\mathbf{F_3}(a,b,c)=(a\lor b)\land(b\lor \lnot c)$ obtained using factors for $\mathbf{F_1}(a,b) = a\lor b$ and $\mathbf{F_2}(b,c)=b\lor \lnot c$ is shown in Table~\ref{tab:factor_ex}(c).

  From the definition of  marginalization and factor product operations, the following lemmas are inferred in the literature.
  \begin{lemma} \label{lem:marg} If $\psi(\mathbf{Y})$ is obtained from $\phi(\mathbf{Y},z)$ after marginalization of the variable $z$, the two factors have the same PF.\end{lemma}
  This is true because after marginalization the partial assignment $\{Y_i\}$ is mapped to the sum of the weights corresponding to $z=0$ and $1$, so that the overall count of all the weights does not change. The lemma extends in a straightforward way if sets of variables are marginalized. This is illustrated in the Tables~\ref{tab:factor_ex}(a) and (d) which have a PF of 3.\\
  \begin{lemma} \label{lem:fp} If $\psi(\mathbf{X},\mathbf{Y}, \mathbf{Z}) = \phi_1(\mathbf{X},\mathbf{Y})\phi_2(\mathbf{Y},\mathbf{Z})$, with $\mathbf{X}\cap \mathbf{Y} \cap \mathbf{Z} = \varnothing$.
  Let $\phi_3(\mathbf{Y})$ denote the factor obtained after marginalization of the set of variables $\mathbf{Z}$. Then
  \begin{equation}
    \mathrm{PF}(\psi) = \mathrm{PF}(\phi_1 \phi_2) = \mathrm{PF}(\phi_1 \phi_3)
  \end{equation}
  \end{lemma}
 The weight for an assignment $Y_i$ in $\phi_3$ is a sum of the weights in $\phi_2$ that contain this assignment. The factor product multiplies the weights in $\phi_1$ that contain $Y_i$ with the corresponding weight in $\phi_2$ or $\phi_3$. The lemma follows since multiplication distributes over addition. Tables~\ref{tab:factor_ex}(c) and (e) correspond to the factor products of $\psi_3(a,b,c)=\phi_1(a, b)\phi_2(b,c)$ and $\psi_5(b,c) = \psi_1(b)\phi_2(b,c)$ respectively and clearly have 5 as their PF.\\
\begin{figure}[b]
     \centering
\begin{tikzpicture}[font=\footnotesize]
\node[circle, draw, inner sep=3] (x1) at (-2, 0) {${V_1}$};
\node[circle, draw, inner sep=3] (x2) at (0,0 ) {${V_2}$};
\node[circle, draw, inner sep=3] (x3) at (2, 0 ) {${V_3}$};

\node at (-2, -0.5) {$\phi_1(\mathbf{X},\mathbf{Y})$};
\node at (0., -0.5) {$\phi_2(\mathbf{Y},\mathbf{Z})$};
\node at (2, -0.5) {$\phi_3(\mathbf{Z},\mathbf{P})$};

\draw[->,very thick] (x3) to node[midway, above, yshift=0.2cm]{$\psi_3(\mathbf{Z})$} (x2);
\draw[->,very thick] (x2) to node[midway, above, yshift=0.2cm]{$\psi_2(\mathbf{Y})$}  (x1);

\draw[->] (1.5,0.2) to (0.5,0.2);
\draw[->] (-0.5,0.2) to (-1.5,0.2);
\end{tikzpicture}

     \caption{Equation \eqref{eq:mp} represented as a message passing algorithm on a tree. $\phi(.)$ are the initial factors assigned to the nodes and $\psi(.)$ are the messages. $V_1$ is the root node}
     \label{fig:mp}
 \end{figure}

  The two properties basically imply that variables present in a single factor can be marginalized before computing the factor product, without affecting the overall PF. This is the basis of the message passing algorithm. \\
  \textbf{Message Passing (MP):} The sum-product MP algorithm uses a combination of the two operations, sum (Marginalization) and product (Factor product), to obtain metrics like PF. Let $\phi_1(\mathbf{X}, \mathbf{Y}), \phi_2(\mathbf{Y,Z})$ and $\phi_3(\mathbf{Z}, \mathbf{P})$ be factors, defined over disjoint sets $\mathbf{X}, \mathbf{Y}, \mathbf{Z}$ and $\mathbf{P}$. If $\xi$ represents their factor product, using lemmas \ref{lem:marg} and \ref{lem:fp}, we obtain
  \begin{align}
  \textrm{PF}(\xi) &=  \sum_{X,Y,Z,P \in \mathbf{X},\mathbf{Y},\mathbf{Z},\mathbf{P}}  \phi_1(X,Y) \phi_2(Y,Z)\phi_3(Z,P)\nonumber \\
  &=  \sum_{X,Y \in \mathbf{X},\mathbf{Y}}  \phi_1(X,Y) \sum_{Z \in \mathbf{Z}} \phi_2(Y,Z)\sum_{P \in \mathbf{P}}\phi_3(Z,P) \nonumber \\
  &=  \sum_{X,Y \in \mathbf{X},\mathbf{Y}}  \phi_1(X,Y) \sum_{Z \in \mathbf{Z}} \phi_2(Y,Z) \psi_3(Z)\nonumber \\
  &= \sum_{X,Y \in \mathbf{X},\mathbf{Y}}  \phi_1(X,Y) \psi_2(Y)
  \label{eq:mp}
\end{align}
The factors $\psi_3(\mathbf{z})$ and $\psi_2(\mathbf{Y})$ are the messages obtained by marginalizing $\mathbf{P}$ and $\mathbf{Z}$ in $\phi_3(\mathbf{Z,P})$ and $\phi_2(\mathbf{Y,Z})$ respectively. The message is combined with the other factors in a (factor) product operation, before being summed over all possible values of $x,y$ to get the PF. This process is illustrated in Fig.~\ref{tab:factor_ex} where the table $\psi_1(b)$ (Fig.~\ref{tab:factor_ex}(d)) is the message that is combined with $\phi_2(a,b)$(Fig.~\ref{tab:factor_ex}(b)) to get the table in Fig.~\ref{tab:factor_ex}(e).

Fig. \ref{fig:mp} shows a graphical representation of equation \eqref{eq:mp}, referred to as the message passing algorithm in the literature.
The graph has the following properties.\\
\textbf{Definition 1: (Join tree)} The graphical representation of computing the PF using marginalization and factor product operations is a rooted directed tree, known as the join or junction tree, in which messages are passed from the leaves to root in a post-order traversal. Each vertex is associated with a set of variables and the initial factors $\phi(\cdot)$ are assigned to vertices that contain the corresponding variables.

Every vertex combines the messages from its children as a factor product and generates a message to the parent using the sum operation.
From equation (\eqref{eq:mp}), it is clear that a variable can only be marginalized after finding the factor product of all factors that contain the variable. To ensure this, the join tree must satisfy a property known as the running intersection property (RIP).\\
\textbf{Property 1: (RIP)}  If a variable $x$ is present in vertices $V_i$ and $V_k$, then every vertex  $V_j$ in the path between $V_i$ and $V_k$ must contain $x$.\\
\textbf{Property 2:} The overall PF can be computed by summing the weights at the root node after finding the factor product of the initial factor assigned to the root node and the messages from the child nodes.

Several join trees are possible for a given set of factors, depending on the order in which the variables are marginalized. The width of a join tree is one less than the maximum number of variables associated with a vertex.\\
\textbf{Property 3:}   The complexity of the message passing algorithm is exponential in the treewidth, defined as the minimum width over all possible join tree representations.

The method described is one possible method for join-tree construction, called the variable elimination technique. There are other methods in the literature. Tree decomposition of a graph has been used in SharpSAT-TD \cite{SharpSAT-TD} and GPUSAT\cite{Fichte19}, but they use it to accelerate computation of a single model count and not for optimization of multiple queries.

 \section{System used and Error Metrics}

Our system consists of the exact and approximate circuits with $n$ input bits  and a subtractor that computes the difference of the two outputs as shown in Figure \ref{fig:miter}. The Tseitin transformation is used to convert the synthesized system to a CNF formula $\mathbf{F}$.
Model counting or \#SAT computes the total number of satisfying solutions for $\mathbf{F}$, which we denote \id{sat-count(\mathbf{F})}.

 Let $\mathbf{Y}, \hat{\mathbf{Y}} \from \B^{n} \to \B^{m}$ denote the outputs of the exact and approximate circuits respectively. $\mathcal{E}= \mathbf{Y} - \hat{\mathbf{Y}}$ is the $m+1$ bit error,  obtained as the output of the subtractor in the twos complement form.
 We denote the $i^{th}$ bit of $\mathbf{Y}, \hat{\mathbf{Y}}$ and $\mathcal{E}$ as $y_{i}, \hat{y}_{i}$ and $ e_{i}$, respectively.
\id{sat-count}($e_i$) denotes the model count of $F$ with $e_i$ set to one i.e.,\id{sat-count}$(\mathbf{F}\mid e_i=1)$.

Error metrics can be computed by obtaining the model count for various assignments of the error bits \cite{Vasicek19}. The ER, MAE, MSE and WCE can be computed as follows.\\
\textbf{Error Rate:}(ER) It is the probability that the output of the approximate circuit is erroneous.

The miter generally used  consists of $m$ XOR gates, with inputs $y_i$ and $\hat{y}_i$, followed by a tree of OR gates \cite{Vasicek19}. Since our system uses a single subtractor as the miter, we compute the ER as one minus probability that the error is zero.
    \begin{align}
    \mathrm{ER} &= 1 - \frac{1}{2^n}\id{sat-count\left(\bigvee_{i=0}^{m} e_i == 0 \right)}
    \label{eq:er}
\end{align}
\textbf{Mean absolute error}:(MAE) It can be computed as
\begin{multline}
	    \mathrm{MAE}(y, \hat{y}) =  \frac{1}{2^n}\biggl( \id{sat-count}(e_{m}) \biggr. \\ \left.+ \sum_{i=0}^{m-1}  2^i\id{sat-count}(e_{i} \oplus e_{m})\right)
     \label{eq:mae}
\end{multline}
\textbf{Mean Squared error}:(MSE) It can be computed as
 \begin{multline}
     \mathrm{MSE}(y, \hat{y}) = \frac{1}{2^{n}} \biggl ( \sum_{i=0}^{m} 2^{2i}\id{sat-count}(e_{i})~   \biggr.\\
     \left. + \sum_{i=0}^{m-1} \sum_{j = i + 1}^{m-1}2^{i+j+1}\id{sat-count}(e_{i} \land e_{j})\right .\\ \left .- \sum_{i=0}^{m-1} 2^{i+m}\id{sat-count}(e_{i} \land e_{k}) \right)
     \label{eq:mse}
\end{multline}
 \textbf{Worst case error}:(WCE) It can be either positive ($e_m=0$) or negative ($e_m=1$). To find positive WCE, the formula used is $\mathbf{F}\land \neg e_m$. Then, starting with the $m^{th}$ bit, $e_{m-1}$, each bit is tested sequentially for SAT. Following the test for $e_i$, a unit clause $e_i$ or $\neg e_i$ is added to the formula before proceeding to the next bit. The positive WCE is the binary number with all the SAT bits set to one and rest to zero. For negative WCE, the same procedure is followed with SAT for complement of the error bits and the formula set to $\mathbf{F}\land e_m$. One is added in the end to the negative WCE to get the two's complement. The overall WCE is the maximum of the magnitudes of positive and negative error.

 In the methods used in the literature, each \id{sat-count} in the above equations corresponds to an output of an error miter. MAE computation, for example, will have $m+1$ miter outputs and will make $m+1$ independent calls to the model counter.

 \begin{figure}[t]
  \centering
\begin{tikzpicture}[scale=0.7, transform shape,
    block/.style={
        draw, rectangle, inner sep=5, align=center, fill=white,
    },
    >=stealth,
]
    \node (inp_text) at (0.75,0) [left] {Inputs ($\mathbf{Z}$)};
    \coordinate (fork) at (1.5,0);
    \node[block] (acc) at (3.5, 1.) {Accurate \\ ($f$)};
    \node[block] (app) at (3.5, -1.) {Approximate \\ ($\hat{f}$)};
    \node[block, inner sep=10] (sub) at (7.5, 0) {Subtractor};
    \node (out_text) at ($(sub.east)+(0.5,0)$) [right, align=left] {Output Error \\ ($\mathcal{E} = \{e_0, \dots, e_m\}$)};
    \draw (inp_text.east) -- (fork);
    \draw[->] (fork) |- (acc.west);
    \draw[->] (fork) |- (app.west);
    \draw[->] (acc.east) -- ++(1.2cm, 0) node[midway, above] {($\mathbf{Y}$)}
          |- ($(sub.west)+(0, 0.25cm)$);
    \draw[->] (app.east) -- ++(0.9cm, 0) node[midway, above] {($\mathbf{\hat{Y}}$)}
          |- ($(sub.west)-(0, 0.25cm)$);
    \draw[->] (sub.east) -- (out_text.west);
\end{tikzpicture}     \caption{System used for computation of error metrics}
     \label{fig:miter}
\end{figure}

\section{Proposed Algorithm}
\begin{figure}[b]
 \centering
\begin{minipage}{0.45\columnwidth}
\begin{align*}
\mathbf{F}=\overbrace{(a\lor b \lor c)}^{v_1} \land \overbrace{(\lnot b \lor c \lor d)}^{v_2}\\
\land  \underbrace{(b\lor \lnot d \lor e )}_{v_3} \land \underbrace{(e \lor \lnot a)}_{v_4}
\end{align*}
\end{minipage}
\begin{minipage}{0.45\columnwidth}
\centering
\begin{tikzpicture}[scale=0.35, vertex/.style={minimum size=0.5pt, draw,circle,color=black, draw, fill=none, inner sep=1.5pt}, font=\footnotesize, thick]
\node[vertex] at (0,0) (v1) {$v_1$};
\node[vertex] at (0,3) (v2) {$v_2$};
\node[vertex] at (5,3) (v3) {$v_3$};
\node[vertex] at (5,0) (v4) {$v_4$};
\draw (v1) -- node[pos=0.5, below]{$a$} (v4);
\draw (v1) -- node[pos=0.5, left]{$c$} (v2);
\draw (v2) -- node[pos=0.5, above]{$d$} (v3);
\draw (v3) -- node[pos=0.5, right]{$e$} (v4);
\node[label=right:$b$, yshift=-1] at (1.5, 1.5) (i) {};
\draw (v1) -- (i.center) -- (v2) (i.center) -- (v3);
\end{tikzpicture}
\end{minipage}
\caption{CNF formula and its equivalent hypergraph}\label{fig:hypergraph}
\end{figure}

\begin{figure*}[tb]
\centering
\begin{tikzpicture}[font=\footnotesize]
\node[anchor=center] at (-4.5,2) {\underline{\small (a) Construct $\mathbf{G}$ and convert to tree}};
\node[circle, draw, inner sep=3] (x1) at (-5.9, -0.75) {${V_1}$};
\node[circle, draw, inner sep=3] (x2) at (-3.1, -0.75) {${V_2}$};
\node[circle, draw, inner sep=3] (x3) at (-4.5, 1.2) {${V_3}$};
\node at (-6, -1.3) {$\{\mathbf{F_1}, \mathbf{X_1}=\{a,b,c\}\}$};
\node at (-2.9, -1.3) {$\{\mathbf{F_2}, \mathbf{X_2}=\{a,d,e_1\}\}$};
\node at (-2.9, 1.25) {$\{\mathbf{F_3}, \mathbf{X_3}=\{b,d,e_2\}\}$};
\draw[color=cyan, very thick] (x1) to node[midway, above, sloped]{\color{black}$\mathbf{X_{1,2}}=\{a\}$} (x2);
\draw (x2) to node[pos=0.45, above, sloped]{$\mathbf{X_{2,3}}=\{d\}$}  (x3) to node[midway, above, sloped]{$\mathbf{X_{1,3}}=\{b\}$}  (x1);

\draw [->, thick] (-4.5, -1.25) --node[pos=0.5, anchor=center, rotate=90, align=center]{Merge $\mathbf{F_1}, \mathbf{F_2}$ and\\ marginalize $\{a,c\}$} ++(0, -2.7);

\node[circle, draw, inner sep=3] (x12) at (-5.9, -4.4) {${V_{1}}$};
\node[circle, draw, inner sep=3] (x33) at (-3.1,-4.4) {${V_3}$};
\draw (x12) to node[pos=0.5, anchor=south]{$\mathbf{X_{1,3}}=\{b,d\}$} (x33);
\draw[dashed, thick, color=gray!80!white] (-1.5,1.5) -- ++(0, -6.8);
\node[align=center] at (-5.9, -5.05) {$\{\mathbf{F_1}\land\mathbf{F_2}$,\\ $\mathbf{X_1}=\{b,d,e_1\}\}$};
\node at (-3.1, -4.9) {$\{\mathbf{F_3}, \mathbf{X_3}=\{b,d,e_2\}\}$};

\node[anchor=center] at (0,1.7) {$\mathbf{T_1(X_1)}$};
\node[anchor=north] (t1) at (0, 1.5)
{
\begin{tabular}{|c|c|c|c|}
\hline
\multicolumn{3}{|c|}{$\mathbf{S(F_1)}$} & \multirow{2}{*}{$\mathbf{c}$} \\ \cline{1-3}
$a$ & $b$ & $c$ & \\ \hline\hline
F & F & F & 1\\ \hline
F & T & F & 1\\ \hline
T & T & F & 1\\ \hline
T & T & T & 1\\ \hline
\end{tabular}
};
\node[anchor=center] at (3,1.7) {$\mathbf{T_2(X_2)}$};
\node[anchor=north] (t2) at (3, 1.5)
{
\begin{tabular}{|c|c|c|c|}
\hline
\multicolumn{3}{|c|}{$\mathbf{S(F_2)}$} & \multirow{2}{*}{$\mathbf{c}$} \\ \cline{1-3}
$a$ & $d$ & $e_1$ & \\ \hline\hline
F & F & T & 1\\ \hline
F & T & F & 1\\ \hline
T & F & T & 1\\ \hline
T & T & T & 1\\ \hline
\end{tabular}
};
\node[anchor=center] at (3,2.2) {\underline{\small (b) Table computations}};
\node[anchor=center] at (6,1.7) {$\mathbf{T_3(X_3)}$};
\node[anchor=north] (t3) at (6, 1.5)
{
\begin{tabular}{|c|c|c|c|}
\hline
\multicolumn{3}{|c|}{$\mathbf{S(F_3)}$} & \multirow{2}{*}{$\mathbf{c}$} \\ \cline{1-3}
$b$ & $d$ & $e_2$ & \\ \hline\hline
F & F & F & 1\\ \hline
F & T & T & 1\\ \hline
T & F & F & 1\\ \hline
T & T & T & 1\\ \hline
\end{tabular}
};
\node[anchor=north] (t12) at (1.5, -1.75)
{
\begin{tabular}{|c|c|c|c|c|c|}
\hline
\multicolumn{5}{|c|}{$\mathbf{S(F_{1}\land F_2)}$} & \multirow{2}{*}{$\mathbf{c}$} \\ \cline{1-5}
$a$ & $b$ & $c$ & $d$ & $e_1$ & \\ \hline\hline
F & F & F & F & T & 1\\ \hline
F & F & F & T & F & 1\\ \hline
\rowcolor[HTML]{BE9F9F}
F & T & F & F & T & 1\\ \hline
F & T & F & T & F & 1\\ \hline
\rowcolor[HTML]{BE9F9F}
T & T & F & F & T & 1\\ \hline
\rowcolor[HTML]{9FBE9F}
T & T & F & T & T & 1\\ \hline
\rowcolor[HTML]{BE9F9F}
T & T & T & F & T & 1\\ \hline
\rowcolor[HTML]{9FBE9F}
T & T & T & T & T & 1\\ \hline
\end{tabular}
};
\node[rectangle, rounded corners, inner sep=1mm, draw] (merge) at (1.5,-1.2) {Merge $\mathbf{T_1(X_1)}$ and $\mathbf{T_2(X_2)}$};
\draw[thick, ->] ($(t1.south)+(0.5,0.12)$) to (merge);
\draw[thick, ->] ($(t2.south)+(-0.5,0.12)$) to (merge);
\draw[thick, ->] (merge) -- (t12);
\node (t12m) at (5.75, -3.8)
{
\begin{tabular}{|c|c|c|c|}
\hline
$b$ & $d$ & $e_1$ & $\mathbf{c}$\\ \hline\hline
F & F & T & 1\\ \hline
F & T & F & 1\\ \hline
\rowcolor[HTML]{BE9F9F}
T & F & T & 3\\ \hline
T & T & F & 1\\ \hline
\rowcolor[HTML]{9FBE9F}
T & T & T & 2\\ \hline
\end{tabular}
};
\node at (6.4,-2.) {$\mathbf{T_1(X_1)}$};
\node[rectangle, rounded corners, inner sep=1mm, draw] (marg) at (5.75,-1.2) {Marginalize $\{a,c\}$};
\draw[thick, ->] (t12) to[out=40, in=180] (marg);
\draw[thick, ->] (marg) -- (t12m);
\draw[dashed, thick, color=gray!80!white] (7.5,2.2) -- ++(0, -7.5);
\node (m123) at (9.5, -3.25)
{
\begin{tabular}{|c|c|c|}
\hline
\multicolumn{2}{|c|}{$\mathbf{m_{1\rightarrow 3}}$} & \multirow{2}{*}{$\mathbf{c}$} \\ \cline{1-2}
$b$ & $d$ & \\ \hline\hline
F & F & 1\\ \hline
F & T & 1\\ \hline
T & F & 3\\ \hline
T & T & 3\\ \hline
\end{tabular}
};
\node[anchor=center] at (9.5,2) {\underline{\small (c)  MP for model count of $\mathbf{F}$}};
\node[circle, draw, inner sep=3] (x1) at (8.5, 1) {${V_3}$};
\node[circle, draw, inner sep=3] (x2) at (8.5,-1 ) {${V_1}$};
\draw[very thick] (x1) to node[midway, right, xshift=0.2cm]{$\mathbf{m_{1\rightarrow 3}}$} (x2);
\draw[->] (8.8,-0.7) to (8.8,0.7);
\node at (9.5, 1.) {$\mathbf{T_3(X_3)}$};
\node at (9.5,-1) {$\mathbf{T_1(X_1)}$};

\node at (9.5, -4.8) {$\sum_{b,d,e_2}\mathbf{T(X_3)m_{1\rightarrow 3}} = 8$};

\end{tikzpicture}
\caption{(a) Graph $\mathbf{G}$ constructed using partitions $\mathbf{F_{1}, F_{2}, F_{3}}$ of the CNF formula $\mathbf{F}$ with error variables $e_1$ and $e_2$ (b) Merging $\mathbf{T_{1}}$ and $\mathbf{T_{2}}$ and marginalization of variables $\{a,c\}$ (c) Message passing}
\label{fig:ce}
\end{figure*}
\begin{algorithm}
\caption{MCAC algorithm} \label{alg:top_level}
\begin{algorithmic}[1]
\Require{CNF $\mathbf{F}$, Number of partitions $P$, Table size threshold $TS$}
\State $\{\mathbf{F}_1, \mathbf{F}_2,\ldots, \mathbf{F}_P\} \gets \proc{Partition}(\mathbf{F}, P)$
\State Construct an undirected-graph $G = (\mathbf{V}, \mathbf{E})$ with\\ $V_i=\{\mathbf{F_i},\mathbf{X_i}, \mathbf{T(F_i)}\}$,
\State $\mathbf{T_i(X_i)}$$\gets$$\proc{AllSATSolutions}(\mathbf{F_i})$$\forall \mathbf{F_i}\in\{\mathbf{F}_1, \mathbf{F}_2,... \mathbf{F}_P\}$
\While{$G$ is \textbf{not} a tree}
\ForAll{$V_i\in V$}
\State $\mathbf{X_{\text{marg}}} \gets \mathbf{X_i} \setminus \bigcup\limits_{E_{ij}\in E} \mathbf{X_j}$ \Comment{variables unique to $\mathbf{X_i}$}
\State \proc{Marginalize}($\mathbf{T(X_i)}$, $\mathbf{X_{\text{marg}}}$)
\EndFor
\State $\mathbf{D}$ $\gets$ \proc{ReduceAndEstimateSize}($G$)
\State Sort edges in increasing order of $\mathbf{D}$
\State L $\gets$ Filter edges in $\mathbf{E}$ with $D_{ij} > TS$ and choose disconnected edges with smallest weights
\ForAll{$E_{ij} = (V_i,V_j)\in L$}
\State $V_i\gets\{\mathbf{F_i}\land \mathbf{F_j}, \mathbf{X_i}\cup \mathbf{X_j}$, \\\proc{FactorProd}$(\mathbf{T(X_i)}, \mathbf{T(X_j)})$\} \Comment{Merge operation}
\EndFor
\State Rebuild $G$ with updated $V_i$
\EndWhile
\For {$metric$ in $error\_metrics$} \Comment{e.g. ER, MAE, MSE}
\ForAll{$e\gets\proc{ErrorAssignments}(metric)$}
  \State $\mathbf{T'(X_i)}\gets\proc{Filter}(\mathbf{T(X_i)}, e)$
  \State $\id{sat-count}(e)\gets\proc{MessagePassing}(G, \mathbf{T'(F_i)})$
\EndFor
\EndFor
\end{algorithmic}
\end{algorithm}
The proposed top-level algorithm is shown in Algorithm~\ref{alg:top_level}.  Firstly, we partition the CNF formula $\mathbf{F}$ using a hypergraph partitioner. For each partition/sub-formula, SAT solver is used to construct a factor (table) comprising all the satisfying solutions for the sub-formula. Following this, we construct a graph $G$, where each vertex corresponds to a partition and its factor.  A series of marginalizations, factor product and conflict resolution operations are performed with the aim of converting $G$ to a rooted directed tree with as few vertices as possible. Following this, the MP algorithm is used to obtain the model count for each setting of the error bits. For each assignment of the error bits $S_e$, we show that the PF at the root node is the required model count (= $\id{sat-count(\mathbf{F}|S_e)}$).
In this section, we discuss the algorithms used in these steps with relevant proofs.
\subsection{Partition $\mathbf{F}$}
\proc{Partition} function in Algorithm~\ref{alg:top_level} involves the following steps: Let $\mathbf{X}$ denote the set of variables in the formula $\mathbf{F}$.
 We first convert $\mathbf{F}$ into a hypergraph using the method in \cite{mann2017}. Each clause in $\mathbf{F}$ is a vertex in the hypergraph and each variable $x\in \mathbf{X}$ is a hyperedge. A hyperedge associated with variable $x$ connects all the vertices(clauses) that contain $x$ or its complement. An example CNF formula with four clauses and its equivalent hypergraph is shown in Fig.~\ref{fig:hypergraph}.

 A hypergraph partitioner~\cite{kahypar23} is then used to partition $\mathbf{F}$ into $P$ partitions, $\mathbf{F}_1, \cdots, \mathbf{F}_P$, such that $\mathbf{F} = \bigwedge_{i=1}^P \mathbf{F}_i$ and $\mathbf{X} = \bigcup_{i=1}^P \mathbf{X_i}$, where $\mathbf{X_i}$ denotes the set of variables in the $i^{th}$ partition. The partitioning objective is to minimize the number of partitions in which a variable is present, with  a constraint on the maximum number of clauses and variables present in a single partition. This effectively limits the number of satisfying solutions in each partition.

\subsection{Graph construction}
 A graph $G(\mathbf{V},\mathbf{E})$ is constructed as follows. The vertices of $G$ are  $V_i=\{\mathbf{F_i},\mathbf{X_i}, \mathbf{T(X_i)}\}$, where $\mathbf{T(X_i)}$ is a table corresponding to the sub-formula $\mathbf{F_i}$, whose construction is explained in the following sub-section. An edge $E_{ij} \in \mathbf{E}$ connects two vertices $V_i$ and $V_j$ in $G$ if $\mathbf{X_{i,j}} = \mathbf{X_i}\cap \mathbf{X_j} \neq \emptyset$. Fig.~\ref{fig:ce}(a) shows the example graph \emph{G} with $\mathbf{X_{i,j}}$ shown on the edges. Note that at this point, \emph{G} is not necessarily a tree.

\subsection{Initial factor construction}
  In this step, we construct a factor  $\phi_i(\mathbf{X}_i): \textrm{Domain}(\mathbf{X}_i) \rightarrow \mathbb{Z}$ associated with each vertex $V_i$, so that
  $\textrm{PF}(\phi_i) =  \id{sat-count}(\mathbf{F_i})$.
  To do this, we first find  the set of all satisfying solutions of  $\mathbf{F_i}$ denoted  $\mathbf{S(F_i)}$ using a \#SAT solver that supports solution enumeration~\cite{cpsat}. This is efficient since the partitioning is done with a limit on the number of clauses and variables in each partition.
  Following this, $X_{ik} \in \textrm{Domain}(\mathbf{X_i})$ is mapped to a count $c(X_{ik})$ as follows.
  \begin{equation}
     \phi_i(X_{ik}) = c(X_{ik}) =
    \begin{cases}
      & 1, \quad X_{ik} \in \mathbf{S(F_i)} \\
      & 0, \quad \textrm{otherwise}
      \end{cases}
  \end{equation}

   Let $\mathbf{T_i(X_i)}$ denote the table representation of the factor $\phi_i$.
  Usually, the number of satisfying solutions is much less than the number of assignments in Domain($\mathbf{X_i})$. Hence, for memory efficiency, we only store the satisfying  assignments i.e., the ones that have a non-zero count, in the table $\mathbf{T_i(\mathbf{X_i})}$. The entries in the initial table are therefore $ \{X_{ik} \in \mathbf{S(F_i)}, c(X_{ik})=1\}$, with $ \id{sat-count}(\mathbf{F}_i)$ = PF($\phi_i) = \sum_{k=1}^n c_{ik} = |\mathbf{S(F_i)}|$.
  Fig.~\ref{fig:ce}(b) shows the initial tables $\mathbf{T_i(\mathbf{X_i})}$ for the graph in Fig.~\ref{fig:ce}(a).

  \subsection{Factor computations: Marginalize, Reduce and Merge}
  The aim is to convert the graph $G$ to a tree. To do this, the factor operations used are  Marginalize, Reduce and Merge.  The operations and implications  of each operation is now described in more detail.\\
 \noindent \textbf{Marginalize:} The marginalization operation is as defined in section \ref{sec:mp}.
 Let $x_m \in \mathbf{X_i}$ denote the variable to be marginalized and $X_{jm}$ denote an assignment of the remaining variables $\mathbf{X_i}\setminus x_m$.
 After marginalization, the count for the entry $X_{jm}$ in the revised table is the number of times $X_{jm}$ appears in the set $\mathbf{S(F_i)}$. Fig. \ref{fig:ce}(b) shows the new counts and entries obtained  after the variables $\{a,c\}$ are marginalized from the table.
 From Lemma \ref{lem:marg} in section \ref{sec:mp}, the PF of a factor does not change after marginalization. Therefore, $ \id{sat-count}(\mathbf{F}_i)$ is the sum of the counts in the revised table.

 In the marginalization step, all variables present in a single factor (partition) are marginalized. From Lemma \ref{lem:fp}, this reduces the size of the tables  without affecting the overall PF.
\emph{Note that since we require the $\id{sat-count}(\mathbf{F}_i|S_e)$ for various settings of the error variables, the error bits are never marginalized. }\\
 \noindent \textbf{Reduce:} The reduce operation, \textit{Reduce}($G(\mathbf{V},\mathbf{E})$) is a conflict resolution mechanism that reduces sizes of tables by removing conflicting assignments of the shared variables in tables associated with adjacent vertices of the graph. Algorithm \ref{alg:reduce} has the procedure.
  Let $\mathbf{T_i(X_i)}$ and $\mathbf{T_j(X_j)}$ be two tables of adjacent vertices with  shared variables $\mathbf{X_{ij}} =  \mathbf{X_i}\cap \mathbf{X_j}$. Let $\mathbf{R_1}(\mathbf{R_2}) = \{S_{ij} \in \textrm{Domain}(\mathbf{X_{ij}})\}$ denote the set of distinct assignments of $\mathbf{X}_{ij}$  present in $\mathbf{T_i(X_i)}(\mathbf{T_j({X}_j)})$. Therefore $\mathbf{R_{12}} = \mathbf{R_1} \cap \mathbf{R_2}$ contains assignments of  $\mathbf{X}_{ij}$ present in both tables. Satisfying assignments
  in $\mathbf{T_i}$ and $\mathbf{T_j}$ that do not contain an entry in the set $\mathbf{R_{12}}$ are deleted.
This is done iteratively. In each iteration, conflicting assignments of shared variables between adjacent vertices are removed from the corresponding tables. The iteration terminates since one of the following two conditions must be satisfied - the conflicting assignments between all adjacent tables are removed, that is, when $|\mathbf{R_1}| = |\mathbf{R_2}| = |\mathbf{R_{12}}|$ in all tables or the formula is unsatisfiable (either one or both table sizes become zero).
\begin{algorithm}[htb]
\caption{Reduce and estimate size} \label{alg:reduce}
\begin{algorithmic}[1]
\Require{Graph $G(\mathbf{V},\mathbf{E})$}
\State flag $\gets$ 0
\State $size\_estimate\gets \emptyset$
\While {flag == 0}
\State flag $\gets$ 1
\ForAll{$E_{ij} \in \mathbf{E}$}
\State $\mathbf{X_{ij} \gets  X_i\cap X_j}$
\State $size\_estimate(E_{ij}) \gets 0$ \Comment{size estimate if $\mathbf{T_i(S_i)}$ and $\mathbf{T_j(S_j)}$ were merged}
\State $\mathbf{R_1} \gets \{S_{ij} \in \textrm{Domain}(\mathbf{X}_{ij})\} \in \mathbf{T_i(X_i)}$
\State $\mathbf{R_2} \gets \{S_{ij} \in \textrm{Domain}(\mathbf{X}_{ij})\} \in \mathbf{T_j(X_j)}$
\State $\mathbf{R_{12}} \gets \mathbf{R_1} \cap \mathbf{R_2}$
\If {$|\mathbf{R_1}| \neq |\mathbf{R_{12}}|$}
\State flag $\gets$ 0
\State Delete entries in $\mathbf{T_i(X_i)}$ with $\mathbf{R_1\setminus R_{12}}$
\EndIf
\If {$|\mathbf{R_2}| \neq |\mathbf{R_{12}}|$}
\State flag $\gets$ 0
\State Delete entries in $\mathbf{T_j(X_j)}$ with $\mathbf{R_2\setminus R_{12}}$
\EndIf
\EndFor
\ForAll {$R_m\in R_{12}$}
\State $c_1 \gets $ Number of entries in $\mathbf{T_i(X_i)}$ with $R_m$
\State $c_2 \gets $ Number of entries in $\mathbf{T_j(X_j)}$ with $R_m$
\State Increment $size\_estimate(E_{ij})$ by $c_1\cdot c_2$
\EndFor
\If {$|\mathbf{T_i(X_i)}| ==0 ~\textrm{or} ~|\mathbf{T_j(X_j)}| ==0$ }
\State Print(UNSAT); flag = 1; break
\EndIf
\EndWhile
\Return{$size\_estimate$}
\end{algorithmic}
\end{algorithm}

In addition to reducing table sizes to retain only pair-wise consistent entries, an estimate of the size of the resultant table on merging $\mathbf{T_i(X_i)}$ and $\mathbf{T_j(X_j)}$. is calculated. For this step, we make the conservative assumption that there is no marginalization possible when these tables are merged. With this assumption, if $\mathbf{T_i(X_i)}$ and $\mathbf{T_j(X_j)}$ were merged, the resultant table contains variables $\mathbf{X_i\cup X_j}=\{\mathbf{X_i\setminus X_{ij},X_{ij},X_j\setminus X_{i,j}}\}$, with $\mathbf{R_{12}}$ as the set of common assignments of $\mathbf{X_{ij}}$. For every entry $R_m\in\mathbf{R_{12}}$, there are as many solutions in the merged table as the product of number of solutions with $R_m$ in $\mathbf{T_i(X_i)}$ and $\mathbf{T_j(X_j)}$. The estimated size of merged table hence is the sum of such products for every $R_m\in\mathbf{R_{12}}$.

  \textit{Reduce}($G(\mathbf{V}, \mathbf{E})$) satisfies the following lemma.
  \begin{lemma}
    On completion of Algorithm \ref{alg:reduce}, if the formula is satisfiable, subsets of variables associated with multiple vertices have consistent assignments in all tables.
  \end{lemma}
  \begin{proof}
    If a subset of variables is present in multiple vertices, the subgraph consisting of these vertices and edges between them is a clique in the graph. There are two iterations in the algorithm that are repeated until termination.
     The inner iteration over all edges of the graph removes conflicting entries
     present in tables of adjacent vertices in the clique. The lemma follows, since the outer iteration ensures that the inner iteration is repeated until all adjacent vertices have consistent tables with respect to shared variables or the formula is
     unsatisfiable.
  \end{proof}

\noindent \textbf{Merge:} Let $V_i$ and  $V_j$ be two adjacent vertices in $G$, with associated tables $\mathbf{T_i(X_i)}$ and $\mathbf{T_j(X_j)}$. \textit{Merge}($V_i, V_j$) merges $V_j$ with $V_i$ by removing $V_j$,  connecting all the neighbours of $V_j$ to $V_i$ and replacing $\mathbf{T_i}$ with the factor product of $\mathbf{T_i}$ and $\mathbf{T_j}$. The variable set associated with $V_i$ after the merge is thus $\mathbf{X_i = X_i\cup X_j}$.  Fig.~\ref{fig:ce}(b) shows an example of merge followed by marginalization.  The tables $\mathbf{T_1(X_1)}$ and $\mathbf{T_2(X_2)}$ are merged, resulting in a table $\mathbf{T_1(X_1\cup X_2)}$. Marginalization of the variables $a$ and $c$ gives the revised set of variables $\mathbf{X_1} = \{b,d,e_1\}$ and the revised table $\mathbf{T_1(X_1)}$ associated with the vertex $V_1$.

  Since the tables have a sparse representation of the factor, the factor product is computed as follows. Let $\mathbf{X_{ij}} =  \mathbf{X_i}\cap \mathbf{X_j}$ denote the variables common to $\mathbf{X_i}$ and $\mathbf{X_j}$ before the merge operation and let $\mathbf{Q} = \mathbf{X_i}\cup \mathbf{X_j}$ denote the variables associated with $V_i$ after the merge operation. $\mathbf{T(Q)}$ denotes the table obtained after the merge operation.
  Let $\{S_k,c(S_k)\} \in  \mathbf{T_i}(\mathbf{X_i})$ and $\{S_l,c(S_l)\} \in \mathbf{T_j}(\mathbf{X_j})$ denote entries in the two tables and $\mathbf{R} = \{S_{ij} \in \textrm{Domain}(\mathbf{X_{ij}})\}$ denote the set of distinct assignments of $\mathbf{X_{ij}}$  present in $\mathbf{T_i}(\mathbf{X_i})$. If we look at the definition of the factor product, it essentially merges entries from the two tables that have the same assignment of shared variables $\mathbf{X_{ij}}$ and multiplies the corresponding weights.
To implement this, entries in $\mathbf{T_i(X_i)}$ are  bucketed, with the elements of the set $\mathbf{R}$ as the bucket index. Let $B_i(R_k)$  denote the bucket corresponding to the index $R_k$.
  The entries of the modified table $T \in \mathbf{T}(\mathbf{Q})$ associated with $V_i$, are obtained as follows
  \begin{align}
    & \forall \{S_{l}\setminus R_k, R_k\} \in \mathbf{T_j(X_j)}, \nonumber \\
    & \quad \forall S_{m}\setminus R_k \in B_i(R_k), \nonumber \\
    & T =  \{ (S_{m}\setminus R_k) \cup R_k  \cup (S_l\setminus R_k), c(S_m)\times c(S_l)\} \nonumber \\
    & \qquad \mathbf{T}(\mathbf{Q}) =  \mathbf{T}(\mathbf{Q}) \cup T
    \label{eq:bucket}
  \end{align}
  Therefore, \textit{Merge}($V_i, V_j$) requires one scan of both tables to obtain all entries of the revised table. Since assignment of shared variables, $R_k$, must be present in both tables, this excludes all entries that will have a zero count in $\mathbf{T}$.  Therefore, the modified table is the ``sparse'' factor product of the two tables, storing only entries that have a non-zero count. The merge operation satisfies the following lemma.
  \begin{lemma} \label{lem:sc}
    The revised table $\mathbf{T(Q)} = \mathbf{T_i(X_i)T_j(X_j)}$ obtained after \textit{Merge}($V_i, V_j$) satisfies
    \begin{equation*}
       \textrm{PF}(\mathbf{T(Q)}) = \id{sat-count}(\mathbf{F_i} \land \mathbf{F_j})
      \end{equation*}
  \end{lemma}
  \begin{proof}
    From equation \eqref{eq:bucket}, each  entry of $\mathbf{T}$ is of the form $T = \{ (S_m\setminus R_k) \cup R_k  \cup (S_l\setminus R_k), c(S_m)\times c(S_l) \}$. If no variable has been marginalized in the two tables,  $T_1 = \{ (S_m\setminus R_k) \cup R_k\} \in \mathbf{S(F_i)} $ and   $T_2 = \{ (S_l\setminus R_k) \cup R_k\} \in \mathbf{S(F_j)} $.
    Since the assignment of shared variables $R_k$ is consistent in $T$,
    it follows that $T \in \mathbf{S(F_i \land F_j)}$. Since $\mathbf{T_i}$ and $\mathbf{T_j}$ contain all satisfying assignments of $\mathbf{F_i}$ and $\mathbf{F_j}$ and the iteration \eqref{eq:bucket} is done over all entries that have a common assignment of shared variables, $\mathbf{T(Q)}$ contains all satisfying entries of $\mathbf{F_i} \land \mathbf{F_j}$. Since no variables have been marginalized, the counts $c(S_m)$ and $c(S_l)$ for all entries remains one and PF($\mathbf{T(Q)}$), which is the sum of all the counts, is \id{sat-count}($\mathbf{F_i} \land \mathbf{F_j}$).

    If some of the variables in either table have been marginalized, $T_1$ and/or $T_2$ are assignments of the unmarginalized variables in the corresponding satisfying set, with the corresponding count increased by the number of times $T_1,T_2$ are present in each set. Once again, since the shared variables of $T_1$ and $T_2$ are consistent, $T$ is an assignment of the unmarginalized variables in $\mathbf{S(F_i \land F_j)}$. From Lemma \ref{lem:fp}, the PF of the table is not affected if variables are marginalized before computing the factor product.
  \end{proof}

  \subsection{Computation of error metrics when \emph{G} is a tree}
  The Reduce, Marginalize and Merge operations are done repeatedly until the graph \emph{G} becomes a tree or the nodes can no longer be merged without the resultant table size exceeding the limit.
   If the operations result in a tree, we assign a root node and convert $G$ to a rooted directed tree on which  message passing (MP) is run to the obtain error metrics. An example rooted directed tree is shown in Fig.~\ref{fig:ce}(c).
  \subsubsection{Filter}
To evaluate any error metric, we require only a subset of the entries in the tables $\mathbf{T_i(X_i)}$. For example, to calculate the ER in \eqref{eq:er}, we require the entries containing error-bits $e_i$ set to 0. For each \id{sat-count}, we create copies of tables with entries that match the required assignment of error bits, $S_e$. $\mathbf{T'_i(X_i)}$ contains the necessary entries from $\mathbf{T_i(X_i)}$ to evaluate \id{sat-count(S_e)}.
  \subsubsection{Message passing}
Algorithm \ref{alg:message} has the main steps in the message passing algorithm on graph $\emph{G}$, which is now a rooted directed tree, with the filtered tables $\mathbf{T'_i(X_{i})}$ associated with vertex $V_i$.
A root vertex $r$ is picked from the vertices. All leaf nodes are added to a queue. A node is popped from the queue and a message is passed from the node to its parent and the corresponding factor product is computed.
A message from vertex $v$ to $u$, $m_{v\rightarrow u}$, is a factor that contains variables $\mathbf{X_u\cap X_v}$. To obtain the message, variables $\mathbf{X_v\setminus X_u}$ are marginalized in $\mathbf{T_v}$ as described in Section~\ref{sec:mp}.
Once the node has received messages from all its children, it is added to the queue. The procedure terminates once we reach the root node, at which point the queue is empty.

\begin{algorithm}
\caption{Message passing on a tree and evaluation of \textit{sat-count}} \label{alg:message}
\begin{algorithmic}[1]
\Require{Directed rooted tree $G\gets (V,E)$}
\State \id{r} $\gets$ Root vertex of $G$
\State $Q$ $\gets$ \{all leaves of $G$\} \Comment{Initialize queue with leaves}
\State $D$ $\gets$ $\emptyset$ \Comment {Processed vertices}
\While{$Q\neq \emptyset$}
\ForAll{$v\in Q$}
\State $u\gets$ parent($v$)
\State $m_{v\to u} \gets \proc{Marginalize}(\mathbf{T_v, X_v \setminus X_u})$
\State $\mathbf{T'_u}(\mathbf{X}_{u}) \gets \mathbf{T'_u}(\mathbf{X}_{v}) \cdot m_{v\to u}$ \Comment{factor-product} \label{li:factor-product}
\EndFor
\State $D\gets D\cup Q$\Comment{Messages passed for all $Q$}
\State $Q\gets \emptyset$
\ForAll{$v\in V$}
\If{$v\notin D$ \textbf{and} children($v$)$\in D$}
\State $Q\gets Q\cup \{v\}$
\EndIf
\EndFor
\EndWhile
\Return $\sum\limits_{s\in \mathbf{S}(r)} c(s)$ \Comment{sum the counts of the root}
\end{algorithmic}
 \end{algorithm}

Recall that the MP algorithm works only if RIP is satisfied. The following theorem proves that if a series of merge and marginalize operations result in a tree structure, then it automatically satisfies RIP.
 \begin{theorem}
		When graph $G$ becomes a tree, (\emph{RIP}) is satisfied.
	\end{theorem}
	\begin{proof}
		The initial graph \emph{G} is constructed so that there is an edge between a  pair of vertices in \emph{G} iff they have variables in common. When the Merge step merges $V_j$ with $V_i$, neighbours of ${V_j}$ are connected to ${V_i}$ with each edge associated with the same common variables. Marginalization does not affect variables associated with the edges. After repeated merge-and-marginalize steps, assume that \emph{G} is converted to a tree. The proof is by contradiction. Assume RIP is not satisfied. This implies there exists ${V_j}$ in the path from $V_i$ to $V_k$ with $x \in \mathbf{X_i}$ and $x \in \mathbf{X_k}$, but $x \notin \mathbf{X_j}$. By definition, there must be an edge from ${V_i}$ to ${V_k}$ resulting in a loop, which is not possible since \emph{G} is a tree.
	\end{proof}

 The required \id{sat-count} can be computed using the following theorem
\begin{theorem}
    On termination of the MP algorithm, the required \id{sat-count} for $\mathbf{F}$ can be obtained as
    \begin{equation*}
     \id{sat-count}(\mathbf{F}) = \sum_{s \in \mathbf{T}(\mathbf{X_r})}c(s)
 \end{equation*}
 where $r$ is the root node and $\mathbf{T}(\mathbf{X_r})$ is the table corresponding to the root node.
\end{theorem}
\begin{proof}
  Each step of of the MP is a marginalization followed by a factor product (Merge). A node receives messages  from its children and computes the factor product with its own table.  The messages are the factors obtained after marginalizing variables in the child that are not present in the parent. Marginalization followed by factor product does not alter the overall PF (Lemma \ref{lem:fp}). The PF of the factor product is the \id{sat-count} of the conjunction of the formulas of the node and its children (Lemma \ref{lem:sc}). The MP algorithm terminates when the root node receives messages from all its children. Therefore, the PF at the root node, given by the sum of all the counts in $\mathbf{T(X_r})$,  is the \id{sat-count} of the formula.
\end{proof}

Fig.~\ref{fig:ce}(c) depicts the message computation, passing,  and factor product for finding the \id{sat-count}($\mathbf{F}$), without setting the error bits. $V_3$ is designated the root node and the factor product of $\mathbf{T}(\mathbf{X_3})$ and the message gives final table at the root, with \id{sat-count}($\mathbf{F}) = 8$.

 We compute ER using equation \eqref{eq:er}. The error bits are set to zero and  the  \id{sat-count} is obtained by running the MP algorithm. For MAE, from equation \eqref{eq:mae} we need the \id{sat-count} of the sign bit $e_m$ and $e_i\oplus e_m$.
In our implementation, instead of adding a  clause for XOR, we compute $\id{sat-count}(e_i=0,e_m=1) + \id{sat-count}(e_i=1,e_m=0)$. This requires $2m$ calls to the message passing algorithm. For MSE computation, we set a pair of relevant error bits to one and find the corresponding \id{sat-count}, as shown in equation \eqref{eq:mse}. This requires $m + m(m+1)/2$ calls to the message passing algorithm.

\subsection{Error metric computation if \emph{G} is not a tree}
If no two tables can be merged without exceeding the threshold, the reduce, merge, marginalize and message passing steps are continued after setting the error bits. Setting the error bits to particular values typically reduces the size of the tables significantly, allowing for further merges.  To compute error metrics, the process has to be repeated for each setting of the error bits. Alternately, if the number of error bits are small, it may be possible to directly compute the probability mass function (PMF) of the output error. The PMF can then be used to compute all the required metrics.

\subsection{Table data structure}

The tables used in our merge algorithm are hash tables with (satisfying solutions, counts) forming the (\textit{key, value}) pairs. The \textit{key} is a 64-bit integer, with each bit representing a variable. Each satisfying solution is thus a 64-bit integer. An example with an 8-bit key is shown in Figure \ref{fig:tds}. The value is the associated count, stored as a double precision number. A runtime efficient hash table~\cite{abseil-cpp} is used in our implementation.
\begin{figure}
\centering
\scriptsize
\begin{minipage}[t]{0.45\columnwidth}
\centering
$\mathbf{T(F)}$\\
\begin{tabular}{|c|c|c|c|}
\hline
\multicolumn{3}{|c|}{$\mathbf{S(F)}$} & \multirow{2}{*}{$\mathbf{count}$} \\ \cline{1-3}
$a$ & $b$ & $c$ & \\ \hline\hline
F & F & F & 1\\ \hline
F & T & F & 2\\ \hline
T & T & F & 1\\ \hline
T & T & T & 4\\ \hline
\end{tabular}
\end{minipage}
\begin{minipage}[t]{0.45\columnwidth}
\centering
{map:variable$\rightarrow$index}\\
\begin{tabular}{|c|c|}
\hline
\textbf{Variable} & \textbf{Bit-index} \\
\hline
\hline
$c$ & 0\\ \hline
$b$ & 1\\ \hline
$a$ & 2\\ \hline
\hline
\multirow{1}{*}{Unused}& 3:7\\ \hline
\end{tabular}
\end{minipage}

\vspace{2mm}
\begin{minipage}[t]{\columnwidth}
\centering
\underline{absl::flat\_hash\_map:integer$\rightarrow$double}\\
\begin{tabular}{|c|c|c|c|c|c|c|c|c|}
\hline
\multicolumn{8}{|c|}{\textbf{Solution}} & \textbf{count} \\
\multicolumn{8}{|c|}{\textbf{(key)}} & \textbf{(value)} \\
\hline
	\hline
        0 & 0& 0 & 0 & 0 & \textbf{0} & \textbf{0} & \textbf{0} & 1 \\ \hline
0 & 0 & 0 & 0 &0 &\textbf{0} & \textbf{1} & \textbf{0} & 2 \\ \hline
0 & 0 & 0 & 0 & 0 & \textbf{1} & \textbf{1} & \textbf{0} & 1 \\ \hline
0 & 0 & 0 & 0 & 0 &\textbf{1} & \textbf{1} & \textbf{1} & 4 \\ \hline
\end{tabular}
\end{minipage}
  \caption{An example showing the table data structure, with an 8 bit integer used as the key. Unused bits are set to zero. }
  \label{fig:tds}
\end{figure}

\section{Results}

All experiments were done on a Intel i7-13700 CPU with 64GB of RAM, running Debian 13.
\subsection{Classification of benchmarks}
The following benchmarks sets have been used for evaluation in the literature.\\
\textbf{EvoApproxLib} \cite{Mrazek17,Mrazek18}: These include a collection of signed and unsigned 8 to 16 bit adders and multipliers.\\
\textbf{BACS} \cite{BACS,BACS_paper}: This set has several types of circuits including 32 bit adders, 8 and 16 bit multipliers and arithmetic circuits like butterfly and multiply-accumulate.\\
\textbf{GeAr} \cite{GEAR_KIT}: A library of approximate 8, 16 and 32 bit adders.\\
\textbf{VACSEM} \cite{Meng24}: This set has large approximate circuits including 128 and 256 bit adders, 16 bits (and above) multipliers and some approximate versions of EPFL benchmarks.\\
\textbf{LPAA and Gaussian filters}: We created a benchmark set consisting of low power approximate adders (LPAA)  including AMA~\cite{AMA}, AXA~\cite{AXA}, and LOA~\cite{LOA} of various lengths and Gaussian-$3\times 3$ blurring filters \cite{filters1}.

Table \ref{tab:cb} has some of the parameters of the benchmarks that affect the complexity of model counting. The treewidth is estimated \textit{with none of the error bits set to some value.} This  could reduce during the computation of various error metrics when error bits are set.
The number of bits at the output of the subtractor is one more than the number of primary outputs (POs). However, it is possible that not all POs are erroneous, which means that some outputs of the subtractor are identically zero. In the table, \#ERB is the number of outputs of the subtractor that are not identically zero. \textit{We refer to it as the number of output error bits.}
If the number of primary inputs (PIs) are small, exhaustive enumeration  is an option.

We chose a collection of benchmarks from various sets including adders of various lengths from GeAr and VACSEM, multipliers from BACS, EvoApprox and VACSEM, Gaussian filters and a few other arithmetic benchmarks (butterfly, x2 and binsqd (BACS)). In most benchmarks, \#ERBs are not very large. Therefore, we also evaluated our method with LPAAs with upto 90 (signed) error bits, making it challenging to evaluate the MSE.

In all cases, a subtractor was added to the Verilog files for approximate and exact circuits from the benchmark set and the entire system  was synthesized with YOSYS~\cite{yosys} and ABC~\cite{abc}, using a library of basic gates.

\begin{table}
    \scriptsize
\begin{tabular}{|c|c|c|c|c|c|}
\hline
Benchmark  & \#Vars & \#Cls  & \# PIs & \#ERB & Estimated TW  \\
\hline
buttfly (BACS) & 39 & 98 & 32 & 16 &  $<$10 \\ \cline{2-6}
ACA\_II\_N32\_Q16 (GeAr)  & 392 & 1362 & 64 & 10 & $<$10 \\ \cline{2-6}
add128\_10 (VACSEM) & 1336 & 3192 & 256 & 11 & 11 \\ \cline{2-6}
ACA\_I\_N32\_Q8 (GeAr) & 718 & 2949 & 64 & 10 & 15 \\ \hline \hline
AXA3-128 (MCAC) & 505 & 1131 & 256 & 64 & $<$10 \\ \cline{2-6}
LOA-128 (MCAC) & 258 & 390 & 256 & 64 & $<$10 \\ \hline \hline
Gauss3x3-AMA2 (MCAC) & 724 & 2080 & 72 &  5 &  30 \\ \cline{2-6}
Gauss3x3-LOA (MCAC) & 647 & 1854 & 72 & 5 & 33 \\ \cline{2-6}
binsqrd (BACS) & 790 & 2291 & 16 & 4 & 41 \\ \cline{2-6}
mult8 (BACS) & 617 & 1786 & 16 & 16 & 45 \\ \cline{2-6}
mult10 (VACSEM) & 787 & 2259 & 20 & 8 &  36 \\ \hline \hline
mult11 (EvoA) & 1105 & 3440 & 22 & 21 & 72 \\ \cline{2-6}
mult12\_s (EvoA) & 1123 & 3533 & 24 & 25 & 53 \\ \cline{2-6}
mult12 (VACSEM) & 844 & 2617 & 24 & 5 & 50\\ \cline{2-6}
mult15\_10 (VACSEM) & 1353 & 4172 & 30 & 4 & 76 \\ \cline{2-6}
mult16\_10 (VACSEM) & 2232 & 6493 & 32 & 5 & 70\\ \hline
\end{tabular}
\caption{Parameters for classification of benchmarks. The table has selected benchmarks from each class. Vars:Variables, Cls:Clauses, PI: Primary inputs, \#ERB: Output error bits and TW:Treewidth.}
\label{tab:cb}
\end{table}

\subsection{Method used for comparison}
BDD construction was not an option for many of the benchmarks. Of the model counters, GANAK ~\cite{ganak} was the winner of the SAT competition in 2025 and a compiled binary is available.
As mentioned in the introduction, a recent work VACSEM \cite{Meng24} also uses GANAK  along with logic simulation. We could not use VACSEM directly as it requires the CNF file in a specific format, Moreover, verification of WCE, MSE and other probabilities was not possible.

\subsection{Evaluation}
To use GANAK, we split our CNF files into multiple files, each augmented with a specific constraint of the error bits corresponding to a particular miter output and a time-out of 1 hour per call. The results of multiple calls were compiled to get the error metrics. Wherever possible, we verified our results with GANAK.
For some LPAAs, we could verify the MSE against analytical expressions~\cite{celia18}. The EvoApprox benchmarks include results of all metrics computed using either BDDs or exhaustive enumeration.

Table \ref{tab:gamc} shows a comparison of the run times for the benchmarks in Table \ref{tab:cb} for single threaded execution. Since  GANAK in invoked multiple times with different CNF formulas for each metric, we report the average time per call, maximum and total times for both MAE and MSE. Benchmarks for which the total execution time is less than a second are excluded. In addition to the error metrics already discussed, we also include the probability of WCE, denoted P(WCE). It is computed  by the MP algorithm after setting the error bits to the WCE.
\begin{table*}
\centering
\setlength{\tabcolsep}{2.5pt}
\scriptsize
\begin{tabular}{|l|l|r|r|r|r|r|r|r|r|r|r|r|r|r|r|r|r|r|}
\hline
\multirow{3}{*}{Type} &\multirow{3}{*}{Benchmark} & \multicolumn{9}{c|}{GANAK runtime (s)} & \multicolumn{7}{c|}{MCAC runtimes (s)} \\ \cline{3-18}
& & \multirow{2}{*}{ER}  & \multirow{2}{*}{P(WCE)}      & \multicolumn{3}{c|}{MAE}      & \multicolumn{3}{c|}{MSE} & \multirow{2}{*}{Total}& \multirow{2}{*}{P+I+M} & \multirow{2}{*}{WCE} & \multirow{2}{*}{P(WCE)} &\multirow{2}{*}{ER} & \multirow{2}{*}{MAE} & \multirow{2}{*}{MSE} & \multirow{2}{*}{Total}    \\ \cline{5-10}
&  &  & & Avg. & Max & Total & Avg & Max & Total & & & & & & & &  \\ \hline
\multirow{9}{*}{LPAA}     & AMA1               &  0.26     &  0.01     &  2.98     &  3.45     &  381.5   &  1.62    &  3.46     &  3469.0  &  3850.7  &  1.7     &  1.308   &  $<$1ms  &  $<$1ms  &  0.009   &  0.173   &  3.2    \\  \cline{2-18}
                          & AMA2               &  1.27     &  0.02     &  2.53     &  3.15     &  323.5   &  1.97    &  3.21     &  4231.3  &  4556.1  &  1.4     &  1.189   &  $<$1ms  &  $<$1ms  &  0.016   &  0.289   &  2.9    \\  \cline{2-18}
                          & AMA3               &  0.21     &  0.01     &  2.51     &  2.93     &  160.9   &  1.94    &  3.3      &  4155.0  &  4316.1  &  1.2     &  1.085   &  $<$1ms  &  $<$1ms  &  0.006   &  0.100   &  2.4    \\  \cline{2-18}
                          & AMA4               &  0.26     &  0.01     &  3.23     &  3.72     &  412.9   &  2.41    &  3.82     &  5177.4  &  5590.6  &  1.9     &  1.468   &  $<$1ms  &  $<$1ms  &  0.067   &  1.084   &  4.6    \\  \cline{2-18}
                          & AMA5               &  0.03     &  0.07     &  0.03     &  0.04     &  1.9     &  0.03    &  0.07     &  62.8    &  64.8    &  1.2     &  0.233   &  0.002   &  $<$1ms  &  0.034   &  0.813   &  2.3    \\  \cline{2-18}
                          & AXA1               &  0.01     &  2.41     &  5.56     &  8.49     &  356.2   &  2.03    &  6.97     &  4356.6  &  4715.2  &  9.7     &  2.581   &  0.009   &  0.005   &  0.425   &  7.208   &  20.0   \\  \cline{2-18}
                          & AXA2               &  0.01     &  0.01     &  2.31     &  2.75     &  145.6   &  0.11    &  2.99     &  237.1   &  382.7   &  1.9     &  0.667   &  $<$1ms  &  $<$1ms  &  0.004   &  0.072   &  2.7    \\  \cline{2-18}
                          & AXA3               &  0.28     &  0.01     &  0.83     &  1.85     &  53.3    &  0.78    &  1.89     &  373.9   &  427.5   &  1.8     &  0.131   &  $<$1ms  &  $<$1ms  &  0.002   &  0.062   &  2.0    \\  \cline{2-18}
                          & LOA                &  0.03     &  0.07     &  0.01     &  0.01     &  0.5     &  0.01    &  0.01     &  17.6    &  18.2    &  9.2     &  0.611   &  $<$1ms  &  $<$1ms  &  0.001   &  0.020   &  9.8    \\  \hline
                          & GAMA1              &  3274.63  &  32.32    &  -        &  $>$3600  &  -       &  -       &  $>$3600  &  -       &  -       &  4.7     &  5.351   &  $<$1ms  &  $<$1ms  &  $<$1ms  &  $<$1ms  &  10.1   \\  \cline{2-18}
                          & GAMA2              &  792.21   &  5.18     &  -        &  $>$3600  &  -       &  -       &  $>$3600  &  -       &  -       &  4.6     &  7.660   &  $<$1ms  &  $<$1ms  &  $<$1ms  &  $<$1ms  &  12.3   \\  \cline{2-18}
Gaussian                  & GAMA3              &  260.14   &  24.3     &  1533.29  &  3128.8   &  7666.45    &  -       &  $>$3600  &  -       &  -       &  3.4     &  6.656   &  $<$1ms  &  $<$1ms  &  $<$1ms  &  $<$1ms  &  10.1   \\  \cline{2-18}
filters                   & GAMA4              &  7.92     &  3.49     &  66.07    &  332.88   &  726.8   &  10.23   &  28.57    &  204.6   &  937.6   &  4.6     &  3.477   &  $<$1ms  &  $<$1ms  &  $<$1ms  &  $<$1ms  &  8.1    \\  \cline{2-18}
built                     & GAMA5              &  75.08    &  34.21    &  -       &  $>$3600   &  -       &  -       &  $>$3600  &  -       &  -       &  2.1     &  6.426   &  $<$1ms  &  $<$1ms  &  $<$1ms  &  $<$1ms  &  8.6    \\  \cline{2-18}
with                      & GAXA1              &  $>$3600  &  16.91    &  -        &  $>$3600  &  -       &  -       &  $>$3600  &  -       &  -       &  4.0     &  5.922   &  $<$1ms  &  $<$1ms  &  $<$1ms  &  $<$1ms  &  9.9    \\  \cline{2-18}
LPAA                      & GAXA2              &  $>$3600  &  12.22    &  -        &  $>$3600  &  -       &  -       &  $>$3600  &  -       &  -       &  14.0    &  7.857   &  $<$1ms  &  $<$1ms  &  $<$1ms  &  $<$1ms  &  21.9   \\  \cline{2-18}
                          & GAXA3              &  277.21   &  5.11     &  -        &  $>$3600  &  -       &  20.03   &  58.38    &   300.52 &  -       &  2.7     &  0.125   &  $<$1ms  &  $<$1ms  &  $<$1ms  &  $<$1ms  &  2.9    \\  \cline{2-18}
                          & GLOA               &  459.37   &  17.14    &  438.91  &  1333.81  &  2194.53   &  -       &  $>$3600  &  -       &  -       &  2.4     &  5.952   &  $<$1ms  &  $<$1ms  &  $<$1ms  &  $<$1ms  &  8.4    \\  \hline
\multirow{2}{*}{Evo}      & mult11             &  57.59    &  0.58     &  125.99   &  174.89   &  2142.0  &  122.03  &  235.36   & 18548.14 &  20748.3 &  92.1    &  21.525  &  $<$1ms  &  $<$1ms  &  0.143   &  0.036   &  113.8  \\  \cline{2-18}
                          & mult12s\_2         &  4.7      &  0.12     &  193.77   &  364.16   &  3487.8  &  95.46   &  368.37   & 16228.14 &  19720.8 &  47.5    &  22.450  &  $<$1ms  &  $<$1ms  &  0.966   &  0.062   &  71.0   \\  \hline
\multirow{3}{*}{BACS}     & mult8              &  2.85     &  0.06     &  6.6      &  11.38    &  138.6   &  3.65    &  7.09     &  240.6   &  382.2   &  2.00    &  1.653   &  $<$1ms  &  $<$1ms  &  $<$1ms  &  $<$1ms  &  3.7    \\  \cline{2-18}
                          & binsqrd            &  3.32     &  5.61     &  6.6      &  9.47     &  46.2    &  12.36   &  94.22    &  123.6   &  178.8   &  2.76    &  0.032   &  $<$1ms  &  $<$1ms  &  $<$1ms  &  $<$1ms  &  2.8    \\  \cline{2-18}
                          & x2                 &  0.22     &  0        &  0.03     &  0.07     &  0.4     &  0.03    &  0.18     &  1.0     &  1.6     &  0.01    &  0.006   &  $<$1ms  &  $<$1ms  &  $<$1ms  &  $<$1ms  &  0.1    \\  \hline
\multirow{2}{*}{GeAr}     & ACA\_II\_N32\_Q16  &  0.41     &  0.02     &  0.5      &  1.47     &  1.5     &  0.5     &  1.48     &  3.0     &  5.0     &  0.85    &  0.102   &  $<$1ms  &  $<$1ms  &  $<$1ms  &  $<$1ms  &  0.9    \\  \cline{2-18}
                          & ACA\_I\_N32\_Q8    &  1.3      &  0.04     &  1.7      &  2.52     &  110.3   &  0.94    &  2.35     &  529.4   &  641.0   &  2.71    &  0.525   &  $<$1ms  &  $<$1ms  &  0.049   &  0.005   &  3.3    \\  \hline
\multirow{4}{*}{VACSEM}   & add128\_10         &  3.14     &  0.92     &  3.53     &  5.35     &  38.8    &  1.43    &  3.13     &  30.0    &  72.9    &  6.19    &  0.063   &  $<$1ms  &  $<$1ms  &  $<$1ms  &  $<$1ms  &  6.3    \\  \cline{2-18}
                          & add128\_18         &  3.57     &  0.75     &  3.03     &  4.51     &  63.7    &  0.84    &  3.84     &  55.5    &  123.6   &  5.26    &  0.115   &  $<$1ms  &  $<$1ms  &  $<$1ms  &  $<$1ms  &  5.4    \\  \cline{2-18}
                          & mult10             &  12.74    &  2.47     &  0.98     &  4.02     &  14.8    &  3.85    &  86.01    &  138.7   &  168.7   &  3.14    &  0.054   &  $<$1ms  &  $<$1ms  &  $<$1ms  &  $<$1ms  &  3.2    \\  \cline{2-18}
                          & mult12             &  304.89   &  15.09    &  23.24    &  51.79    &  116.2   &  9.29    &  51.72    &  139.4   &  575.6   &  19.14   &  0.050   &  $<$1ms  &  $<$1ms  &  $<$1ms  &  $<$1ms  &  19.2   \\  \hline
\hline
\multirow{2}{*}{VACSEM}  & mult15\_10         &   \multicolumn{8}{c|}{}   &  1720.5 &  \multicolumn{6}{c|}{}    &  246.0      \\  \cline{2-18}
                          & mult16\_10         &   \multicolumn{8}{c|}{}   & 9130.9  &  \multicolumn{6}{c|}{}    &  1397.0     \\  \hline

\end{tabular}
\caption{Comparison of GANAK and MCAC for single-threaded execution. Avg. and Max are the average and maximum runtimes over all  formulas needed for the metric. P+I+M is the runtime for partitioning, initialization and merge. Metrics for mult15\_10 and mult16\_10 were evaluated using relevant assignments of the output error bits. }
\label{tab:gamc}
\end{table*}

In these benchmarks, MCAC spends a majority of runtime in the overhead part, consisting of partitioning, initialising tables and merging the graph until it becomes a tree (P+I+M).
In majority of the cases, the resulting tree has only one or two nodes. This makes it relatively trivial to compute all error metrics, even if there are a large number of MP calls. This is particularly seen in the LPAA benchmarks. Since these benchmarks have low treewidth, each call of GANAK also takes only a few seconds. The thousands of seconds required by GANAK is due to the large number of calls to compute the MSE.
For the Gaussian filters, we could not obtain all results using GANAK in many cases and exhaustive enumeration is not an option. GANAK gave P(WCE) in all cases, which matched with our results. We did an approximate comparison of MAE and MSE  with results obtained using $10^9$ inputs.  MCAC computed all metrics for these filters in around 10s. The two multipliers from EvoApprox take longer than the 10 and 12 bit multipliers from VACSEM. As seen from Table II, both are larger circuits with a higher treewidth and \#ERB than the 12 bit multiplier in the VACSEM benchmarks. For 8-12 bit multipliers,  MCAC runtime is 2-3 orders of magnitude faster than making multiple calls to GANAK.

 There are 21 and 24 instances of 15 and 16 bit multipliers (mult15 and mult16) benchmarks in VACSEM respectively with \#ERB ranging from 1 to 7. In these benchmarks, merge and marginalize did not result in a tree. However, since \#ERB is not large, we were able to compute the metrics by running our algorithm for all possible assignments of the error bits, other than the all zero assignment which is not needed. Using this method we were able to solve 17 out of 21 instances of the mult15 benchmarks with a geomean runtime of 388s. For mult16, we could solve 15 out of 24 instances with geomean runtime of 1798s. The runtime comparison with GANAK for one of the instances in each set is included in Table \ref{tab:gamc}. It shows a 5-6x speedup.

In Table \ref{tab:LPAA}, we report the runtimes to compute the error metrics for three 128-bit low power approximate adders for different \#ERB. It is seen that the runtimes is only weakly dependent on \#ERB. This is because the adders have a low treewidth and the final number of vertices in the tree is very small (sometimes just 1), making message passing trivial.
\begin{table}\centering
\resizebox{\columnwidth}{!}{%
\begin{tabular}{|c|r|c|c|c|c|c|r|r|}
\hline
\multirow{3}{*}{LPAA} & \multirow{3}{*}{\#ERB} & \multicolumn{7}{c|}{Runtime(s)} \\ \cline{3-9}
                         &                  &\multicolumn{2}{c|}{Overhead}         & \multirow{2}{*}{WCE}       & \multirow{2}{*}{ER}        & \multirow{2}{*}{MSE}       & \multirow{2}{*}{MAE}       & \multirow{2}{*}{Total} \\\cline{3-4}
                         &                  & P   & \makecell{I+M}    & & & & &  \\ \hline
\hline

\multirow{3}{*}{AMA3} & 32 & 0.1 & 0.80 & 0.06 & 0.0001 &  0.01 & 0.00 &  1.01 \\ \cline{2-9}
                      & 64 & 0.6 & 1.73 & 0.22 & 0.0001 &  0.16 & 0.01 &  2.72 \\ \cline{2-9}
                      & 90 & 1.2 & 2.66 & 0.67 & 0.0006 &  6.02 & 0.24 & 10.86 \\ \hline

\multirow{3}{*}{AMA4} & 32 & 1.0 & 0.77 & 0.07 & 0.0002 &  0.28 & 0.04 &  2.18 \\ \cline{2-9}
                      & 64 & 1.5 & 1.66 & 0.22 & 0.0004 &  1.54 & 0.08 &  4.99 \\ \cline{2-9}
                      & 90 & 2.0 & 2.38 & 0.31 & 0.0001 &  0.49 & 0.02 &  5.24 \\ \hline

\multirow{3}{*}{AXA2} & 32 & 0.6 & 0.26 & 0.12 & 0.0000 &  0.01 & 0.00 &  1.03 \\ \cline{2-9}
                      & 64 & 0.6 & 0.50 & 0.25 & 0.0000 &  0.06 & 0.00 &  1.41 \\ \cline{2-9}
                      & 90 & 0.7 & 0.68 & 0.37 & 0.0001 &  0.16 & 0.01 &  1.89 \\ \hline
\end{tabular}
}
\caption{Runtimes to evaluate the error metrics for various 128-bit LPAA adders. ERB is the number of erroneous output bits; P and I+M are the runtimes for the partitioner, table initialization and merge.}\label{tab:LPAA}
\end{table}

Table~\ref{tab:filters} has the results for $3 \times 3$ approximate Gaussian blurring filters used in image processing.  These have been recently used for design space exploration in \cite{Pedram23}.
Only LOA has been used in \cite{Pedram23} and they have used an approximate technique to compute the MSE. Using MCAC, we computed the exact MSE for the 3$\times$ 3 Gaussian blurring filters, using various approximate adders. The table shows that the MSE of the approximate filter is a good indicator of the PSNR of blurred images. The runtimes are of the order of 10s, showing that our tool is useful for design exploration. The table shows that use of AMA1, AMA2, AMA5 and LOA leads to better performance than the other adders, with AXA3 giving the worst performance.

\begin{table}\centering
\resizebox{\columnwidth}{!}{%
\begin{tabular}{|c|r|r|c|c|c|}
\hline
LPAA & Runtime &  MSE  &  PSNR & PSNR     & PSNR     \\
     &  \multicolumn{1}{c|}{(s)} &       &       & Lena (B) & Lena (O) \\ \hline
AMA1 & 10.10   &  5.63 & 40.63 & 38.84    & 35.36    \\ \hline
AMA2 & 12.27   &  4.27 & 41.80 & 39.15    & 35.73    \\ \hline
AMA3 & 10.07   & 10.46 & 37.93 & 36.87    & 34.80    \\ \hline
AMA4 &  8.06   & 10.07 & 38.10 & 36.33    & 34.25    \\ \hline
AMA5 &  8.56   &  6.04 & 40.32 & 38.65    & 35.50    \\ \hline
AXA1 &  9.93   &  6.96 & 39.70 & 38.32    & 35.40    \\ \hline
AXA2 & 21.86   & 12.23 & 37.25 & 36.40    & 34.30    \\ \hline
AXA3 &  2.86   & 55.43 & 30.69 & 30.57    & 30.13    \\ \hline
LOA  &  8.35   &  6.78 & 39.81 & 39.12    & 35.70    \\ \hline

\end{tabular}
}
\caption{MSE and the corresponding PSNR of 3$\times$3 Gaussian blurring filters. PSNR-Lena is the PSNR of the approximately blurred Lena image with respect to exact blurring (B) and the original image without blurring (O). }\label{tab:filters}
\end{table}

Fig.~\ref{fig:hist_mult8} shows the histogram of error probabilities for the mult8 benchmark. It was obtained by setting the error bits for each possible value of the error and  finding the resultant \id{sat-count} using the message passing algorithm.
\begin{figure}
\centering
\begin{tikzpicture}
\pgfplotsset{scaled y ticks=false}
\pgfplotsset{every axis/.append style={
line width=0.8pt,
tick style={line width=0.6pt}}}
\begin{axis}[
xmin=-518, xmax=512, ymin=0, ymax=0.02,
xlabel=error, ylabel=P(error), height=3.8cm, width=\columnwidth, font=\small,
xtick={-518,-256,0, 256, 512},
ytick={0, 0.005, 0.010, 0.015, 0.020},
yticklabels={0, 0.005, 0.010, 0.015, 0.020},
grid=major]
\addplot[x=err, y=prob, blue,ybar,fill, bar width=0.01] table {histogram_mult8.txt};
  \end{axis}
\end{tikzpicture}
\caption{Histogram of error probabilities for the BACS benchmark mult8 generated using our algorithm.}\label{fig:hist_mult8}
\end{figure}

\subsection{Discussion and Comparison with related work}
The two works in which multiple metrics have been computed for a large number and variety of instances are \cite{Vasicek19} and \cite{Meng24}. Both are miter based approaches that compute the \id{sat-count} for each miter output independently. In this sense, their methods are similar to making multiple calls to GANAK in our evaluation.
In \cite{Vasicek19}, they were able to solve for ER, MAE and MSE of 32 bits adders and 12 bit multipliers using BDDS, but not using a \#SAT counter. GANAK is able to solve larger instances, but is significantly slower than MCAC.
Interestingly, their runtimes for the 32 bit adder had a strong dependence on the WCE. In MCAC, we have not found such a strong dependence on WCE for even 128 bit adders with a large number of error bits.  This is because the complexity of MCAC depends on the treewidth, which is quite small for adders. As a result, the merge is effective and results in a very small tree, making MP efficient. The effect of treewidth is apparent for mult11 (EvoA) benchmark, for which a larger treewidth leads to a longer runtime, as seen from Tables \ref{tab:cb} and \ref{tab:gamc}. The 12 bit multipliers from EvoA and VACSEM have similar treewidths, but the longer runtime for mult12\_s is due to the larger \#ERB. WCE depends both on the position of the error bit and the \id{sat-count} of the error bit. For these circuits, if \id{sat-count} is small, then using the Reduce algorithm with merge and filtering  before message passing is quite effective in reducing table sizes and improving runtimes, even if it is a higher order error bit.

Both VACSEM  and MCAC have specific advantages and disadvantages. VACSEM has shorter runtimes for ER and MAE, but requires a significantly larger synthesis effort since each miter output of every metric has to be separately synthesized and optimized. This also makes it difficult when the \#ERB is large. MCAC has longer (but competitive) runtimes, but requires minimal synthesis time and effort. It computes all five metrics with a single call. VACSEM also requires a tight coupling between the circuit structure and the CNF formula and does not work for any CNF input.  We could not invoke the logic simulation part of VACSEM with the CNF formulas that we generated for GANAK. MCAC accepts any CNF formula. There is clearly a trade-off in the runtime and the synthesis and logic optimization time required.

Our method of partitioning and merging is better at exploiting variables that can be marginalized than standard methods of tree decomposition, like the one based on variable elimination. However, occasionally, there are some instances that unexpectedly take a larger amount of time due to unbalanced table sizes. This happens if there are a large number of error bits, since we do not marginalize the error bits. For example, LOA which takes about 10s (Table III) runs in less than a second if the partition size is decreased. For these cases, it might be useful to integrate a  standard tree-decomposition method within the framework and use it if the treewidth is small.

\section{Conclusion}
We have proposed an algorithm based on \#SAT and message passing that can be used for exact computation of a variety of error metrics. Besides the standard metrics, we can obtain various probabilities including the entire PMF. We have been able to compute error metrics, including MSE, of several large instances of filters and multipliers.

Our aim was to use a single miter and algorithm for all benchmarks. Although largely successful, the performance could possibly be improved by integrating some of the techniques used in SOTA model counters.  Many parts of the algorithm, like Reduce and error metric computation are also trivially parallelizable. This will significantly improve runtimes, especially for cases where the number of error bits is large. The combination of a high treewidth, low conflicts (which means higher \id{sat-counts}) and a large number of output error bits is challenging for all model counters, including our solver.

\section{Acknowledgments}
We have not used any generative AI tools in the development of the code or in the preparation of the manuscript.

\bibliographystyle{ieeetr}

\bibliography{ref}

@STRING{TCAD = "{IEEE} Trans. Comput.-Aided Design Integr. Circuits Syst."}

@STRING{TCASI = "{IEEE} Trans. Circuits Syst. {I}"}

@INPROCEEDINGS{Yu16,
  author={Yu, Cunxi and Ciesielski, Maciej},
  booktitle={IEEE Computer Society Annual Symposium on VLSI (ISVLSI)}, 
  title={{Analyzing Imprecise Adders Using BDDs -- A Case Study}}, 
  year={2016},
  volume={},
  number={},
  pages={152-157},
  doi={10.1109/ISVLSI.2016.85}}

@INPROCEEDINGS{Venkatesan11,
  author={Venkatesan, Rangharajan and Agarwal, Amit and Roy, Kaushik and Raghunathan, Anand},
  booktitle={2011 IEEE/ACM International Conference on Computer-Aided Design (ICCAD)}, 
  title={MACACO: Modeling and analysis of circuits for approximate computing}, 
  year={2011},
  volume={},
  number={},
  pages={667-673},
  doi={10.1109/ICCAD.2011.6105401}}

@INPROCEEDINGS{Dreshler18,
  author={Froehlich, Saman and Große, Daniel and Drechsler, Rolf},
  booktitle={2018 Design, Automation \& Test in Europe Conference \& Exhibition (DATE)}, 
  title={Approximate hardware generation using symbolic computer algebra employing grobner basis}, 
  year={2018},
  volume={},
  number={},
  pages={889-892},
  doi={10.23919/DATE.2018.8342133}}

@INPROCEEDINGS{Soeken16,
  author={Soeken, Mathias and Große, Daniel and Chandrasekharan, Arun and Drechsler, Rolf},
  booktitle={2016 21st Asia and South Pacific Design Automation Conference (ASP-DAC)}, 
  title={BDD minimization for approximate computing}, 
  year={2016},
  volume={},
  number={},
  pages={474-479},
  doi={10.1109/ASPDAC.2016.7428057}}

@INPROCEEDINGS{Dreshler19,
  author={Froehlich, Saman and Große, Daniel and Drechsler, Rolf},
  booktitle={Design, Automation \& Test in Europe Conference \& Exhibition (DATE)}, 
  title={{One Method - All Error-Metrics: A Three-Stage Approach for Error-Metric Evaluation in Approximate Computing}}, 
  year={2019},
  volume={},
  number={},
  pages={284-287},
  doi={10.23919/DATE.2019.8715138}}

@article{Vasicek19,
author = {Vasicek, Zdenek},
year = {2019},
month = {12},
pages = {1-1},
title = {{Formal Methods for Exact Analysis of Approximate Circuits}},
volume = {PP},
journal = {IEEE Access},
doi = {10.1109/ACCESS.2019.2958605}
}

@INPROCEEDINGS{Ceska17,
  author={Češka, Milan and Matyaš, Jiří and Mrazek, Vojtech and Sekanina, Lukas and Vasicek, Zdenek and Vojnar, Tomas},
  booktitle={IEEE/ACM International Conference on Computer-Aided Design (ICCAD)}, 
  title={Approximating complex arithmetic circuits with formal error guarantees: 32-bit multipliers accomplished}, 
  year={2017},
  volume={},
  number={},
  pages={416-423},
  doi={10.1109/ICCAD.2017.8203807}}

@article{Mrazek18,
author = {Mrazek, Vojtech and Vasicek, Zdenek and Hrbacek, Radek},
title = {Role of circuit representation in evolutionary design of energy-efficient approximate circuits},
journal = {IET Computers \& Digital Techniques},
volume = {12},
number = {4},
pages = {139-149},
doi = {https://doi.org/10.1049/iet-cdt.2017.0188},
url = {https://ietresearch.onlinelibrary.wiley.com/doi/abs/10.1049/iet-cdt.2017.0188},
eprint = {https://ietresearch.onlinelibrary.wiley.com/doi/pdf/10.1049/iet-cdt.2017.0188},
year = {2018}
}

@ARTICLE{Mrazek18tvlsi,
  author={Mrazek, Vojtech and Vasicek, Zdenek and Sekanina, Lukas and Jiang, Honglan and Han, Jie},
  journal={IEEE Transactions on Very Large Scale Integration (VLSI) Systems}, 
  title={Scalable Construction of Approximate Multipliers With Formally Guaranteed Worst Case Error}, 
  year={2018},
  volume={26},
  number={11},
  pages={2572-2576},
  doi={10.1109/TVLSI.2018.2856362}}

@INPROCEEDINGS{ceska22,
  author={Češka, Milan and Matyáš, Jiří and Mrazek, Vojtech and Vojnar, Tomáš},
  booktitle={2022 25th Euromicro Conference on Digital System Design (DSD)}, 
  title={Designing Approximate Arithmetic Circuits with Combined Error Constraints}, 
  year={2022},
  volume={},
  number={},
  pages={785-792},
  doi={10.1109/DSD57027.2022.00110}}

@INPROCEEDINGS{Mrazek22,
  author={Mrazek, Vojtech},
  booktitle={IEEE Computer Society Annual Symposium on VLSI (ISVLSI)}, 
  title={{Optimization of BDD-based Approximation Error Metrics Calculations}}, 
  year={2022},
  volume={},
  number={},
  pages={86-91},
  doi={10.1109/ISVLSI54635.2022.00028}}

@ARTICLE{Keszocze22,
  author={Keszocze, Oliver},
  journal={IEEE Access}, 
  title={BDD-Based Error Metric Analysis, Computation and Optimization}, 
  year={2022},
  volume={10},
  number={},
  pages={14013-14028},
  doi={10.1109/ACCESS.2022.3140557}}

@INPROCEEDINGS{Mrazek25,
  author={Mrazek, Vojtech and Vasicek, Zdenek},
  booktitle={2025 IEEE International Symposium on Circuits and Systems (ISCAS)}, 
  title={AxMED: Formal Analysis and Automated Design of Approximate Median Filters using BDDs}, 
  year={2025},
  volume={},
  number={},
  pages={1-5},
  doi={10.1109/ISCAS56072.2025.11043775}}

@ARTICLE{Qureshi19,
  author={Qureshi, Amina and Hasan, Osman},
  journal={IEEE Transactions on Computer-Aided Design of Integrated Circuits and Systems}, 
  title={Formal Probabilistic Analysis of Low Latency Approximate Adders}, 
  year={2019},
  volume={38},
  number={1},
  pages={177-189},
  doi={10.1109/TCAD.2018.2803622}}

@ARTICLE{Pedram23,
  author={Vaeztourshizi, Marzieh and Pedram, Massoud},
  journal={IEEE Transactions on Very Large Scale Integration (VLSI) Systems}, 
  title={Efficient Error Estimation for High-Level Design Space Exploration of Approximate Computing Systems}, 
  year={2023},
  volume={31},
  number={7},
  pages={917-930},
  doi={10.1109/TVLSI.2023.3273478}}

@INPROCEEDINGS{Rezaalipour23,
  author={Rezaalipour, Morteza and Ferretti, Lorenzo and Scarabottolo, Ilaria and Constantinides, George A. and Pozzi, Laura},
  booktitle={2023 53rd Annual IEEE/IFIP International Conference on Dependable Systems and Networks Workshops (DSN-W)}, 
  title={Multi-Metric SMT-Based Evaluation of Worst-Case-Error for Approximate Circuits}, 
  year={2023},
  volume={},
  number={},
  pages={199-202},
  doi={10.1109/DSN-W58399.2023.00055}}

@INPROCEEDINGS{Meng24,
  author={Meng, Chang and Wang, Hanyu and Mai, Yuqi and Qian, Weikang and De Micheli, Giovanni},
  booktitle={Design, Automation \& Test in Europe Conference \& Exhibition (DATE)}, 
  title={{VACSEM: Verifying Average Errors in Approximate Circuits Using Simulation-Enhanced Model Counting}}, 
  year={2024},
  volume={},
  number={},
  pages={1-6},
  doi={10.23919/DATE58400.2024.10546819}}

@book{Koller2009,
	title={Probabilistic graphical models: principles and techniques},
	author={Koller, Daphne and Friedman, Nir},
	year={2009},
	publisher={MIT press}
}

@article{kahypar23,
author = {{S}chlag, Sebastian and {H}euer, Tobias and Gottesb\"{u}ren, Lars and Akhremtsev, Yaroslav and Schulz, Christian and Sanders, Peter},
title = {{H}igh-{Q}uality {H}ypergraph {P}artitioning},
year = {2023},
issue_date = {December 2022},
publisher = {Association for Computing Machinery},
address = {New York, NY, USA},
volume = {27},
issn = {1084-6654},
url = {https://doi.org/10.1145/3529090},
doi = {10.1145/3529090},
journal = {ACM J. Exp. Algorithmics},
month = {Feb},
articleno = {1.9},
numpages = {39}
}

@article{mann2017,
author = {Mann, Zolt\'{a}n \'{A}d\'{a}m and Papp, P\'{a}l Andr\'{a}s},
title = {{Guiding SAT Solving by Formula Partitioning}},
journal = {International Journal on Artificial Intelligence Tools},
volume = {26},
number = {04},
year = {2017},
doi = {10.1142/S0218213017500117},
URL = {https://doi.org/10.1142/S0218213017500117}
}

@ARTICLE{AMA, 
author={Gupta, Vaibhav and Mohapatra, Debabrata and Raghunathan, Anand and Roy, Kaushik},  journal=TCAD,   title={{Low-Power Digital Signal Processing Using Approximate Adders}},  
year={2013}, 
volume={32},  
number={1}, 
pages={124-137}, 
doi={10.1109/TCAD.2012.2217962}}

@INPROCEEDINGS{filters1,  
author={de Oliveira, Julio and Soares, Leonardo and Costa, Eduardo and Bampi, Sergio},  booktitle={ IEEE 7th Latin American Symposium on Circuits   Systems (LASCAS)},   title={{Exploiting approximate adder circuits for power-efficient Gaussian and Gradient filters for Canny edge detector algorithm}}, 
year={2016}, 
volume={},
number={}, 
pages={379-382}, 
doi={10.1109/LASCAS.2016.7451089}}

@INPROCEEDINGS{Venkataramani12,
  author={Venkataramani, Swagath and Sabne, Amit and Kozhikkottu, Vivek and Roy, Kaushik and Raghunathan, Anand},
  booktitle={DAC Design Automation Conference 2012}, 
  title={SALSA: Systematic logic synthesis of approximate circuits}, 
  year={2012},
  volume={},
  number={},
  pages={796-801},
  doi={10.1145/2228360.2228504}}

@ARTICLE{LOA,  
author={Mahdiani, H. R. and Ahmadi, A. and Fakhraie, S. M. and Lucas, C.},  
journal=TCASI,   
title={{Bio-Inspired Imprecise Computational Blocks for Efficient VLSI Implementation of Soft-Computing Applications}}, 
year={2010}, 
volume={57}, 
number={4},  
pages={850-862}, 
doi={10.1109/TCSI.2009.2027626}}

@INPROCEEDINGS{AXA,  
author={Yang, Zhixi and Jain, Ajaypat and Liang, Jinghang and Han, Jie and Lombardi, Fabrizio},  
booktitle={IEEE International Conference on Nanotechnology (IEEE-NANO 2013)},
title={{Approximate XOR/XNOR-based adders for inexact computing}},   
year={2013}, 
volume={}, 
number={}, 
pages={690-693}, 
doi={10.1109/NANO.2013.6720793}}

@INPROCEEDINGS{GEAR_KIT,
author={Shafique, Muhammad and Ahmad, Waqas and Hafiz, Rehan and Henkel, Jörg},
  booktitle={2015 52nd ACM/EDAC/IEEE Design Automation Conference (DAC)}, 
  title={A low latency generic accuracy configurable adder}, 
  year={2015},
  volume={},
  number={},
  pages={1-6},
  doi={10.1145/2744769.2744778}}

@misc{BACS,
author = {{Brown University Scale Lab}}, 
title= {{BACS: Benchmarks for approximate circuit synthesis}}, 
howpublished= "https://github.com/scale-lab/BACS",
year={2023} 
}

@ARTICLE{BACS_paper,
  author={Scarabottolo, Ilaria and Ansaloni, Giovanni and Constantinides, George A. and Pozzi, Laura and Reda, Sherief},
  journal={Proceedings of the IEEE}, 
  title={{Approximate Logic Synthesis: A Survey}}, 
  year={2020},
  volume={108},
  number={12},
  pages={2195-2213},
  doi={10.1109/JPROC.2020.3014430}}

@INPROCEEDINGS{Mrazek17,
  author={Mrazek, Vojtech and Hrbacek, Radek and Vasicek, Zdenek and Sekanina, Lukas},
  booktitle={Design, Automation \& Test in Europe Conference \& Exhibition (DATE), 2017}, 
  title={EvoApprox8b:Library of Approximate Adders and Multipliers for Circuit Design and Benchmarking of Approximation Methods}, 
  year={2017},
  volume={},
  number={},
  pages={258-261},
  doi={10.23919/DATE.2017.7926993}}

@inproceedings{ganak,
title={{GANAK: A Scalable Probabilistic Exact Model Counter}},
author={Sharma, Shubham and  Roy, Subhajit and  Soos, Mate and  Meel, Kuldeep S.},
booktitle={Proceedings of International Joint Conference on Artificial Intelligence (IJCAI)},
month={8},
year={2019}
}

@inproceedings{buddy,
author={J. Lind-Nielsen and H. Cohen},
title={BuDDy—A Binary Decision Diagram Package},
note={https://sourceforge.net/projects/buddy/},
}

@InProceedings{SharpSAT,
author="Thurley, Marc",
editor="Biere, Armin
and Gomes, Carla P.",
title="sharpSAT -- Counting Models with Advanced Component Caching and Implicit BCP",
booktitle="Theory and Applications of Satisfiability Testing - SAT 2006",
year="2006",
publisher="Springer Berlin Heidelberg",
address="Berlin, Heidelberg",
pages="424--429",
isbn="978-3-540-37207-3"
}

@INPROCEEDINGS{celia18,
  author={Celia, D. and Vasudevan, Vinita and Chandrachoodan, Nitin},
  booktitle={IEEE International Symposium on Circuits and Systems (ISCAS)}, 
  title={{Probabilistic Error Modeling for Two-part Segmented Approximate Adders}}, 
  year={2018},
  volume={},
  number={},
  pages={1-5},
  doi={10.1109/ISCAS.2018.8351273}}

@misc{cudd,
  title = {{CUDD}},
  author = {Fabio Somenzi},
  note = {https://github.com/ssoelvsten/cudd},
url = {https://github.com/ssoelvsten/cudd},
}

@inproceedings{Mishchenko06,
author = {Mishchenko, Alan and Chatterjee, Satrajit and Brayton, Robert and Een, Niklas},
title = {Improvements to combinational equivalence checking},
year = {2006},
isbn = {1595933891},
publisher = {Association for Computing Machinery},
address = {New York, NY, USA},
url = {https://doi.org/10.1145/1233501.1233679},
doi = {10.1145/1233501.1233679},
booktitle = {Proceedings of the 2006 IEEE/ACM International Conference on Computer-Aided Design},
pages = {836–843},
numpages = {8},
location = {San Jose, California},
series = {ICCAD '06}
}

@misc{abc,
  title = {{ABC}},
  note = {https://github.com/berkeley-abc/abc}
}

@InProceedings{Fichte19,
author="Fichte, Johannes K.
and Hecher, Markus
and Zisser, Markus",
editor="Schiex, Thomas
and de Givry, Simon",
title="An Improved GPU-Based SAT Model Counter",
booktitle="Principles and Practice of Constraint Programming",
year="2019",
publisher="Springer International Publishing",
address="Cham",
pages="491--509",
isbn="978-3-030-30048-7"
}

@inproceedings{D4,
author = {Lagniez, Jean-Marie and Marquis, Pierre},
title = {An improved decision-DNNF compiler},
year = {2017},
isbn = {9780999241103},
publisher = {AAAI Press},
booktitle = {Proceedings of the 26th International Joint Conference on Artificial Intelligence},
pages = {667–673},
numpages = {7},
location = {Melbourne, Australia},
series = {IJCAI'17}
}

@InProceedings{SharpSAT-TD,
  author =	{Korhonen, Tuukka and J\"{a}rvisalo, Matti},
  title =	{{Integrating Tree Decompositions into Decision Heuristics of Propositional Model Counters}},
  booktitle =	{27th International Conference on Principles and Practice of Constraint Programming (CP 2021)},
  pages =	{8:1--8:11},
  series =	{Leibniz International Proceedings in Informatics (LIPIcs)},
  ISBN =	{978-3-95977-211-2},
  ISSN =	{1868-8969},
  year =	{2021},
  volume =	{210},
  editor =	{Michel, Laurent D.},
  publisher =	{Schloss Dagstuhl -- Leibniz-Zentrum f{\"u}r Informatik},
  address =	{Dagstuhl, Germany},
  URL =		{https://drops.dagstuhl.de/entities/document/10.4230/LIPIcs.CP.2021.8},
  URN =		{urn:nbn:de:0030-drops-152992},
  doi =		{10.4230/LIPIcs.CP.2021.8}
}

@article{C2D,
  title={On probabilistic inference by weighted model counting},
  author={Mark Chavira and Adnan Darwiche},
  journal={Artif. Intell.},
  year={2008},
  volume={172},
  pages={772-799},
}

@article{Jensen90,
title = "Bayesian updating in recursive graphical models by local computation",
author = "Finn Jensen and Steffen Lauritzen and Kristian Olsen",
year = "1990",
language = "English",
volume = "4",
pages = "269--282",
journal = "Computational Statistics Quarterly",
}

@misc{cpsat,
  title = {{CP-SAT}},
  version = { v9.12 },
  author = {Laurent Perron and Frédéric Didier},
  organization = {Google},
  url = {https://developers.google.com/optimization/cp/cp_solver/},
  note = {https://developers.google.com/optimization/cp/cp\_solver/},
  date = { 2025-02-17 }
}

@software{abseil-cpp,
  title = {{Abseil - C++ Common Libraries}},
  year = {2018},
  publisher = {GitHub},
  url = {https://github.com/abseil/abseil-cpp},
  note = {https://github.com/abseil/abseil-cpp}
}

@misc{yosys,
  author = {Clifford Wolf and others},
  title = {{Yosys Open SYnthesis Suite}},
  year = {2026},
  publisher = {GitHub},
  journal = {GitHub repository},
  note = {https://github.com/YosysHQ/yosys}
}

@inproceedings{pearl1982,
  title={Reverend Bayes on inference engines: A distributed hierarchical approach},
  author={Pearl, Judea},
  booktitle={Proceedings of the National Conference on Artificial Intelligence (AAAI-82)},
  pages={133--136},
  year={1982},
  organization={AAAI}
}

\end{document}